# Dynamic patterns of knowledge flows across technological domains: empirical results and link prediction

Jieun Kim[1*] and Christopher L. Magee[1,2]

[1]Institute for Data, Systems and Society, Massachusetts Institute of Technology, 77 Massachusetts Avenue, Cambridge, MA 02139, USA

[2]International Design Center, Massachusetts Institute of Technology 77 Massachusetts Avenue, Cambridge, MA 02139, USA

Email: Jieun Kim (kimjieun@mit.edu) and Christopher L. Magee (cmagee@mit.edu)

*Corresponding author: Jieun Kim

**Abstract**

The purpose of this study is to investigate the structure and evolution of knowledge spillovers across technological domains. Specifically, dynamic patterns of knowledge flow among 29 technological domains, measured by patent citations for eight distinct periods, are identified and link prediction is tested for capability for forecasting the evolution in these cross-domain patent networks. The overall success of the predictions using the Katz metric implies that there is a tendency to generate increased knowledge flows mostly within the set of previously linked technological domains. This study contributes to innovation studies by characterizing the structural change and evolutionary behaviors in dynamic technology networks and by offering the basis for predicting the emergence of future technological knowledge flows.

# 1. Introduction

Technological evolution is typically shaped by problem solving activity which integrates knowledge from the same and/or different technology areas, leveraging the cumulative character of knowledge. While there is a tendency to generate knowledge flows within a given technology, there are also important knowledge spillovers[1] across technologies and these have the potential to increase technological variety (Schoenmakers and Duysters, 2010). This study focus on the latter case, the interactions across technological domains. Novelty of individual inventions (Kim *et al.*, 2016), importance of individual inventions (Schoenmakers and Duysters, 2010) and possibly convergence of technologies (Caviggioli, 2016) are all important topics that are related to knowledge flow across domains.

Previous studies have attempted to analyze technological knowledge flows in patent networks based on patent citation information (Acemoglu, Akcigit, and Kerr, 2016; Ko, Yoon, and Seo, 2014; Érdi *et al.*, 2013; Lee and Kim, 2010; Chang, Lai, and Chang, 2009; Han and Park, 2006; Jaffe, Trajtenberg, and Henderson, 1993). Their typical unit of analysis has been not only individual patents (Érdi *et al.*, 2013; Chang *et al.*, 2009), but the class of the patent office data such as IPC (International Patent Classification) (Acemoglu *et al.*, 2016), USPC (US Patent Classification) (Ko *et al.*, 2014), Derwent Class Code indexed by technology experts of Thomson Reuters (Luan, Liu, and Wang, 2013). However, ambiguity can arise as the concept of "a technology" and "an industry" are often if not usually conflated whereas technologies as technically understood cut across industries (Benson and Magee, 2015). This research is aimed at understanding non-economic knowledge flows among technologies, thus a carefully defined unit of analysis is utilized.

Although significant effort has been made to understand the evolution of technology networks (Funk and Owen-Smith, 2016; Kim, Cho, and Kim, 2014; No and Park, 2010; Shin and Park, 2010), studies typically construct networks based on posteriori or historical patterns of evolution. In the analysis of dynamic trends of knowledge flows, there is a lack of work attempting to predict a priori patent network structures. As a remedy, in an attempt to understand and predict dynamic evolution of technological knowledge networks, this study uses link prediction methodology, which estimates the likelihood of the existence of a link between two nodes in the future based on observed links and the attributes of nodes (Lü and Zhou, 2011; Liben-Nowell and Kleinberg, 2007).

---

[1] We differentiate economic spillovers between sectors from (technological) knowledge spillovers between technological domains.

The key questions in this study are as follows: how strong are the technological knowledge flows for specific technology domains with other specific domains? and how do those cross-domain links change over time? Can we predict the future cross-domain links in a backcasting experiment? To address these questions, this paper documents dynamic patterns of technological knowledge flow networks, based on measuring patent citations in 29 technological domains (TDs) defined by Magee *et al.* (2016). Dynamic patterns are identified and predicted in the network of knowledge flows between TDs. The paper explores the empirical knowledge flow pattern in the period 1976-2013, as derived from the United States Patent and Trademark Office (USPTO) patent citation data.

The remainder of this paper is organized as follows: Section 2 discusses essential prior work on technological knowledge flow networks and link prediction methodology. Section 3 presents the process for and results of constructing cross-TD networks and identifying dynamic patterns of changes in cross-TD links. Section 4 implements experiments using link prediction methods to test their effectiveness as predictors of cross-TD network evolution. Section 5 integrates the results from previous sections and discusses implications. The final section draws conclusions, highlights limitations and makes suggestions for future research.

## 2. Background

*2.1. Technological knowledge networks*

The notion that technological innovation is the result of the recombination of existing components is deeply rooted in the literature on the history of technological change (Ruttan, 1959; Usher, 1954). Thus, one can often describe inventions as a combination of prior technologies. The inherent combinatorial characteristic of innovation has led scholars to focus on the question of how new technologies build on prior art and on how inventors combine and transfer knowledge across technological domains. Specifically, the mechanism of analogical transfer (Basnet and Magee, 2016; Weisberg, 2006; Gentner and Markman, 1997) has been shown to apply quite broadly and is the most important cognitive mechanism used by inventors. Such recombination of ideas underlies ongoing technological change and is one aspect of generating economic spillover or benefits to a given sector not due to that sectors efforts alone. In particular, the analogical transfer of ideas to new domains means that part of an inventor's original idea necessarily spills over to other firms and other sectors generating positive externalities (the so-called 'knowledge spillovers') for the economy.

Previous work on knowledge spillovers has exploited the comprehensive information provided by patent data to examine how knowledge flows from one invention to the other. Specifically, a patent citation is a reference to prior art for legal purposes and as such represents a proportion of knowledge used in the citing patent that originated from or was already disclosed by the cited patent. Jaffe *et al.* (2000) reported the analysis of R&D manager surveys suggesting that although patent citations carry a fair amount of noise, they a provide a reasonably good indication of a 'learning trail', representing the knowledge transfer process. As a result, the aggregate citation flows within and between technology fields, sectors, geographic areas, etc. can be used as proxies for knowledge flow intensity (Schoenmakers and Duysters, 2010; Verspagen and De Loo, 1999; Karki, 1997; Jaffe *et al.*, 1993; Trajtenberg, 1990). Furthermore, Clough *et al.* (2015) have shown that the level of redundancy—measured by transitive reduction—in the edges of patent citation network is much smaller than in that of academic paper citation network. This suggests that patent citations are better representations of technological knowledge flow than scientific publication citations are of scientific knowledge flows.

Beyond the seminal contributions to knowledge flows by patent citation analysis by Trajtenberg (1990) and Jaffe *et al.* (1993), abundant empirical studies using patent citations to measure knowledge flows have been carried out at the level of an indiviual (Alcácer and Gittelman, 2006; Jaffe *et al.*, 2000), at the firm level (Cho, Kim, and Kim, 2015; Duguet and MacGarvie, 2005), and at an industry level (Park, Lee, and Park, 2009; Han and Park, 2006) and at the national level (Chen and Guan, 2016; Shih and Chang, 2009). Patent citation information thus has been widely used to analyze linkages among patents in a given technology, linkages between technologies, persistence of technological influence and the impact of new patents, as well as the structure of knowledge networks between industries or nations. Studies using main path analysis on patent citation networks for specific technology fields also demonstrate their usefulness for mapping the knowledge and technology trajectory of the real world (Park and Magee, 2017; Nomaler and Verspagen, 2016; Martinelli and Nomaler, 2014; Choi and Park, 2009; Mina *et al.*, 2007; Verspagen, 2007). Network science and social network analysis on the patent citation networks have been widely used to capture the overall structure of patents and technologies including the complicated interactions in technological evolution (Lee, Lee, and Sohn, 2016; Cho *et al.*, 2015; Choi and Hwang, 2014; Kim *et al.*, 2014; Érdi *et al.*, 2013; Cho and Shih, 2011). Other studies focused on the optimal network structures for knowledge diffusion (Shin and Park, 2010). However, prior attempts that analyse patent citation networks have focused only on the observed networks and thereby, they have been limited to an ex post analysis and do not establish potentially predictive models.

An important issue regarding all of the work on paths within or between technologies concerns the unit of analysis. If one is interested only at the patent level or a individual inventor level or even at the firm or national level, the unit of analysis is clear. However, most studies also attempt to study technologies and technological fields but spend little time showing an objective connection to technologies as they exist. An exception is the work of Park and Magee (2017) who utilize a tighter definition but also operationally link a gathered set of patents to technological artifacts whose performance can be measured. Since our focus is on cross-technological flow of knowledge, it is important that we have a clear unit of analysis. Thus, we follow the approach of Park and Magee and describe the definition of technological domains and the methodology for obtaining patent sets to represent such domains in more detail in section 3.1.

*2.2. Link prediction*

2.2.1. Fundamentals

Link prediction attempts to predict the emergence of future links in complex networks based on the available information, such as observed links and nodes' attributes (Lü and Zhou, 2011; Liben-Nowell and Kleinberg, 2007). It not only predicts potential links but also missing and spurious links (Guimerà and Sales-Pardo, 2009). Link prediction is broadly applied in various fields such as biological, social, and information systems where nodes represent biological elements like proteins and genes, individuals, computers, web users, and so on (Lü and Zhou, 2011). In biological networks, accurate predictors can be applied to seek the most promising latent links, which is much less costly than blindly checking all possible interaction connections (Guimerà and Sales-Pardo, 2009; Clauset, Moore, and Newman, 2008). Link prediction has also been used in the analysis of social networks, such as the prediction of the collaborations in co-authorship networks (Liben-Nowell and Kleinberg, 2007), the estimation of collaborative influence (Perez-Cervantes *et al.*, 2013), and the detection of the relationships between community users (Valverde-Rebaza and de Andrade Lopes, 2013). In information systems, link prediction can serve as a significant technique in information retrieval, such as the prediction of words, topics or documents in Wikipedia (Itakura *et al.*, 2011), and in recommender systems, such as e-commerce recommendations (Li *et al.*, 2014) and friend recommendation (Esslimani, Brun, and Boyer, 2011). Moreover, the link prediction approaches can be applied to solve the classification problem in partially labeled networks, such as the detection of anomalous email (Huang and Zeng, 2006), and differentiating fraudulent and legitimate users in cell phone networks (Dasgupta, Singh, and Viswanathan, 2008).

The problem of link prediction and structural definition is as follows. Consider a network $G=(V, E)$, where $V$ is the set of nodes and $E$ is the set of edges connecting nodes. For two nodes $u, v \in V$, $e=(u,v) \in E$ represents an interaction between $u$ and $v$. Given time periods, $t = 1, \ldots, T$, an evolving network is defined as $G_t = (V_t, E_t)$, where $E_t$ is the set of undirected edges, whether new or recurring, between nodes in $V_t$ within time stamp $t$. In this study, $V$ is the set of technological domains (TDs) and $V_t$ can differ according to their occurrence within $t$; $E_t$ represents the knowledge flows or spillovers among TDs within time period $t$. In this setting, the link prediction problem can be formulated as follows: Estimate the likelihood of a potential link in $E_{t+1}$, between two nodes in $V_t$. Link prediction involves the choice of a predictor, a function or algorithm that calculates a likelihood score for the existence of a link between two nodes in the next time period. Network topology-based structural similarity metrics are generally used as predictors (Zhu and Xia, 2016; Valverde-Rebaza and de Andrade Lopes, 2013; Liben-Nowell and Kleinberg, 2007). Applying a predictor to the training network at $t$ yields a number of predictions. In practice, the prediction step results in a list of potential links with an associated likelihood score *Sim*. By ranking the potential links in decreasing order of *Sim* and choosing a threshold, one can obtain a predicted network at *t*+1.

2.2.2. Topology-based predictors

In this study, 8 similarity metrics are tested as predictors of future associations among TDs. According to their characteristics, these predictors can be subdivided into two broad categories: (1) local or neighborhood-based metrics; and (2) global metrics. In all the definitions, *u*, *v*, *z* denote nodes in the network and similarities are always evaluated between two different nodes. *Γ(u)* denotes the set of nodes which are neighbors of node *u* in the network- that is all nodes that are directly linked to *u*.

    Local predictors are solely based on the neighborhoods of the two nodes. Many networks have a natural tendency towards *triadic closure*: if two links a-b and b-c exist, there is a tendency to form the closure a-c (Bianconi *et al.*, 2014; Guns, 2014). This property is closely related to assortativity and and was empirically confirmed in social networks (Bianconi *et al.*, 2014; Kleinberg, 2008) and collaboration networks (Ter Wal, 2014; Liben-Nowell and Kleinberg, 2007). For technological knowledge flow, it assumes that if a given domain shares useful knowledge with two other domains, then those two domains should be able to share knowledge with each other. The networks we consider have the relative strength of knowledge flows as the weights of links and we use weighted similarity scores (Zhu and Xia, 2016; Murata and Moriyasu, 2007) defined below with $w(u,v)$ denoting the weight of the link between nodes *u* and *v*.

(a) *CommonNeighbors* (Newman, 2001) is the most widespread, basic and simplest type of metric that measures the number of nodes with which two adjacent nodes have a direct association. It is known to perform well when a network is highly clustered (Lü and Zhou,

2011). The weighted CommonNeighbors is calculated as:

$$Sim(u,v) = \sum_{z \in \Gamma(u) \cap \Gamma(v)} w(u,z) + w(v,z) \quad (1)$$

(b) *Jaccard* predictor (Salton and McGill, 1983), which is commonly used in information retrieval, considers the probability that two nodes have common neighbors. It is introduced to normalize the effect of neighborhood size (the number of total neighbors of two nodes) in the CommonNeighbors metric: If both *u* and *v* have many neighbors, they are automatically more likely to have more neighbors in common. The Jaccard coefficient for weigted networks can be extended as:

$$Sim(u,v) = \sum_{z \in \Gamma(u) \cap \Gamma(v)} \frac{w(u,z) + w(v,z)}{\sum_{p \in \Gamma(u)} w(p,u) + \sum_{q \in \Gamma(v)} w(q,v)} \quad (2)$$

(c) *Adamic–Adar* predictor (Adamic and Adar, 2003) starts from the hypothesis that a '*rare*' (i.e., low-degree) neighbor is more likely to indicate a future connection than a high-degree one. In many cases, rare features are more telling; documents that share the phrase "for example" are probably less similar than documents that share the phrase "clustering coefficient". In social networks, an unpopular person (someone with not a lot of friends) may be more likely to introduce a particular pair of his friends to each other. In the case of predicting technological domain linkages, the conceptual basis is similar making the assumption that a domain with few connections to other domains is more likely to be connected to domains that will connect in the future. Thus, this metric considers both the common neighbors and the common neighbors' neighbors. Two nodes are likely to connect in the future, if they have more nodes in common and the common nodes have a smaller number of neighbors. The Adamic-Adar measure for weighted networks is:

$$Sim(u,v) = \sum_{z \in \Gamma(u) \cap \Gamma(v)} \frac{w(u,z) + w(v,z)}{\log(1 + \sum_{r \in \Gamma(z)} w(z,r))} \quad (3)$$

(d) *ResourceAllocation* (Zhou, Lu, and Zhang, 2009) is motivated by the resource allocation dynamics on complex networks. Consider a pair of nodes, *u* and *v*, which are not directly connected. The node *u* can send some resource to *v*, with their common neighbors playing the role of transmitters. In the simplest case, we assume that each transmitter has a unit of resource, and will equally distribute it to all its neighbors. The weighted ResourceAllocation between *u* and *v* can be defined as the amount of resource *v* received from *u*, which is:

$$Sim(u,v) = \sum_{z \in \Gamma(u) \cap \Gamma(v)} \frac{w(u,z) + w(v,z)}{\sum_{r \in \Gamma(z)} w(z,r)} \quad (4)$$

It is based on a hypothesis similar to that of Adamic-Adar, but yields a slightly different ranking: the ResourceAllocation index punishes the high-degree common neighbors more heavily than Adamic-Adar.

(e) *PreferentialAttachment* (Barabási *et al.*, 2002; Price, 1976) is based on the idea of '*the rich gets richer*', or '*power-law*'; users with many friends tend to create more connections in the future. It can be shown that the product of the degrees of nodes *u* and *v* is proportional to the probability of a link between *u* and *v*. Thus, the weighted PreferentialAttachment is also known as the degree product and computed as:

$$Sim(u,v) = \sum_{p \in \Gamma(u)} w(p,u) * \sum_{q \in \Gamma(v)} w(q,v) \quad (5)$$

This metric is related to small-world networks, which were illustrated through "six degrees of separation" in social networks by Travers and Milgram (1967). It is broadly known that many knowledge networks, beyond social networks, exhibit the small-world property of having short average path lengths between any two nodes, despite being highly clustered (Watts and Strogatz, 1998). Thus it is known to perform well when a network has small, disconnected clusters (Lü and Zhou, 2011). Examples include scientific collaborations in mathematics and neuro-science (Barabási *et al.*, 2002) and the patent citation network in radio frequency identification (Hung and Wang, 2010). Likewise, we conjecture that the evolution of cross-TD networks might follow PreferentialAttachment because it is consistent with the oft-used concept of general purpose technologies (Moser, Nicholas, and Nicholas, 2015; Jovanovic and Rousseau, 2005). It may be noted that this index does not require any node neighbor information; therefore, it has the lowest computational complexity of our metrics.

Global predictors are the methods based on the ensemble of all paths between two nodes. These predictors recognize that, even if two nodes do not share any common neighbors, they still may be related and form a link in a later time period. In the case of technological knowledge flow, use of these metrics represents an assumption that knowledge flows occur along identifiable paths through the network and this is consistent with the concept of trajectories (Dosi, 1982). For each of the three metrics discussed below, the fundamental assumption of why they might work is that TDs which are on a common trajectory are more likely to cite each other than those not on such a trajectory. A straightforward measure of relatedness is path distance. Likewise, random walk based on transition probabilities from a node to its neighbors can be used to denote the destination from a current node.

(a) The *Katz* metric (Katz, 1953) is a variant of the shortest path distance and is also based on the ensemble of all paths. Let *A* denote the (full) adjacency matrix of the network. The element $a_{uv}$ is the weight of a link between nodes *u* and *v* or 0 if no link is present. Each element $a_{uv}^{(k)}$ of $A^k$ (the k-th power of *A*) has a value equal to the number of walks, i.e. the set of all paths, with length *k* from *u* and *v*. Thus, the Katz metric directly sums the number of all walks that exist. However, as longer walks usually indicate a weaker association between the start and end node, it introduces a free (damping) parameter $\beta$ (0 $< β < 1$), representing the 'probability of effectiveness of a single link'. Thus, each walk with length $k$ has a probability of effectiveness $β^k$, as shown in equation 6. For links between TDs the underlying hypothesis is that more and shorter walks between two TDs indicate a stronger relatedness. It is known to perform well when a network has long average distance (Lü and Zhou, 2011).

$$Sim(u,v) = \sum_{k=1}^{\infty} β^k a_{uv}^{(k)} \qquad (6)$$

(b) *RootedPageRank* is a modification of PageRank (Brin *et al.*, 1998), a core algorithm used by search engines to rank search results. Assume the existence of a random walker, who starts at a random node, randomly chooses one of its neighbors and navigates to that neighbor, again randomly chooses a neighbor and so on. Moreover, at every node, there is a small chance that the walker is 'teleported' to a random other node in the network. The 'chance of advancing to a neighbor' is $α$ ($0 < α < 1$) and the chance of teleportation is $1-α$. For link prediction purposes, the random walk assumption of the original PageRank is altered by the not randomized teleportation: the walker is always teleported back to the same root node. The proximity score between node pairs $u$ and $v$ is calculated in this method as follows:

$$Sim(u,v) = -H_{u,v} \cdot π_v \qquad (7)$$

where $H_{u,v}$ is the hitting time or the expected number of steps required for a random walk from $u$ to reach $v$ and $π_v$ is the stationary distribution weight of $v$ under the following random walk condition: (a) with probability $α$ that returns to $u$, or (b) with probability $1-α$ of jumping to a random neighbor of the current node.

(c) *SimRank* (Jeh and Widom, 2002) is a measure of the similarity between two nodes in a network. The SimRank hypothesis can be summarized as: *nodes that link to similar nodes are similar themselves*. It begins with the assumption that any node is maximally similar to itself: $sim(u,u)=1$, Then it employs a 'decay factor', $\gamma$ ($0 < \gamma < 1$), to determine how quickly similarities or weights of the connected nodes decrease as they get farther away from the original nodes. SimRank can also be interpreted by the random walk process, that is, it measures how soon two random walkers, respectively starting from nodes $u$ and $v$, are expected to meet at a certain node.

$$Sim(u,v) = \frac{\gamma}{|\Gamma(u)| \cdot |\Gamma(v)|} \sum_{p \in \Gamma(u)} \sum_{q \in \Gamma(v)} Sim(p,q) \tag{8}$$

## 3. Cross-TD networks empirically characterized

### 3.1. Data and methods

The unit of analysis in this studiy is a technological domain (TD), defined by Magee *et al.* (2016) as "the set of artifacts that fulfill a specific generic function utilizing a particular, recognizable body of knowledge". This definition essentially decomposes generic functions along the lines of established bodies of knowledge. Thus the domains are connected not only to the economy, but also to science and other technical knowledge. The generic functional classification is described in terms of operands (information, energy, and material) being changed by operations (storage, transportation, and transformation). When each domain defined in this manner is linked operationally to a set of patents, it is able to be connected to non-patent information such as the rate of improvement of artifacts that represent the domain (Triulzi, Alstott, and Magee, 2017; Benson and Magee, 2015). The analysis in the current paper covers patents for 29 TDs found using the classification overlap method (28 from Benson and Magee 2013, 2015), one from Basnet (2015) as shown in Table 1. Despite the fact that these patent sets have been shown to be more complete and relevant than sets usually studied, there are patents in each set (5-25%) that may not be good representatives of the domain. Thus, some "random" cross domain citations are likely and in the work below, we focus on linkages that are greater than random expectation (we call them strong links) partly to avoid this noise source but also because we are interested in a significant level of knowledge flow between domains.

Table 1. The 29 TDs in the generic function format (Magee *et al.*, 2016; Koh and Magee, 2006) with abbreviations used throughout this paper

| Operation | Operand | | |
|---|---|---|---|
| | Information | Energy | Material |
| Storage | ・Semiconductor information storage (SIS)<br>・Magnetic information storage (MIS)<br>・Optical information storage (OIS) | ・Electrochemical batteries (BAT)<br>・Capacitors (CAP)<br>・Flywheel (FLY)<br>・Permanent magnetic materials (PMM) | |
| Transportation | ・Electrical telecommunication (ET)<br>・Optical telecommunication (OT)<br>・Wireless telecommunication (WT) | ・Electrical power transmission (EPT)<br>・Superconductivity (SCD) | ・Aircraft transport (AIR) |
| Transformation | ・Integrated circuit processors (IC)<br>・Electronic computation (EC)<br>・Camera sensitivity (CAM)<br>・Magnetic resonance imaging (MRI)<br>・Computerized tomography scan (CT)<br>・Genome sequencing (GS) | ・Combustion engines (CE)<br>・Electrical motors (EM)<br>・Solar photovoltaic power (SPP)<br>・Wind turbines (WIND)<br>・Fuel cells (FC)<br>・Incandescent lighting (IL)<br>・LED lighting (LED) | ・Milling machines (MIL)<br>・3D printing (3D)<br>・Photolithography (PLG) |

We started by collecting the number of citations between the patents of the TDs annually from 1976 to 2013. The overall number of collected patents for the 29 TDs are 502,444; among the total 7,074,439 patent-to-patent citations, the number of specified inter-domain citations are 290,059 (4%), the number of intra-domain (self) citations are 2,117,399 (30%), and others (~66%) are citations to undefined domains beyond the 29 listed in Table 1.

Exploring annual citation patterns, the time window is set as four or five years, which yields 8 periods for 38 years (from 1976 to 2013) as follows: T1 as 1976~1979, T2 as 1980~1983, T3 as 1984~1988, T4 as 1989~1993, T5 as 1994~1998, T6 as 1999~2003, T7 as 2004~2008, and T8 as 2009~2013. Second, the citation matrix, which we treat as undirected, is constructed using the number of citations between the specified TDs. We standardize all cells in this matrix by calculating *citation scores (CS)* from equation (9):

$$cs_{ijt} = \frac{c_{ijt}/(p_{it}p_{jt})}{c_t/p_t^2} \qquad (9)$$

where *c* is the number of citations, *p* is the number of patents, *t* is the periods from T1 to T8, the subscripts *i* and *j* indicate TDs, and the absence of a subscript indicates an aggregation of TDs. The equation expresses the number of citations between a pair of TDs relative to its expected value, if citations were completely random (Nomaler and Verspagen, 2016) for each time period. Thus, a CS value larger than 1 implies that the citation frequency between these domains is more than for a random occurrence given the citation and patent numbers at that time period. Thus, we call the links whose CS is larger than 1 *strong links*, and those between 0 and 1 *weak links*. For each of the 8 periods, a matrix including 406 non-self citation cells is constructed. As a final step, cross-TD networks[2] are constructed using citation matrices for the 8 periods: the nodes are the 29 TDs and the links are connected with weight CS from equation 9.

## 3.2. Distribution of cross-TD links

Among the 406 non-self citation potential linkages, 57 cells turn out to be zero in every time period. 51 cells are more than one for at least one period whereas 298 cells have values ranging from >0 to 1 for all time periods. See Table 2 for basic statistics in each period.

---

[2] See Appendix 1.

The average of CS is always less than 1, i.e., random expectation, consistent with the maximum values being far greater than 1. As shown in the number of links according to CS range in Table 2 and in CS boxplots (Figure 1), most cross-domain links are weaker than expected by random occurrence in every period and the max values are the definite outliers. Moreover, citations to other patents significantly outnumber the citations involved in even the strongest inter-domain linkages. For example for CT-MRI with a CS of 13.42 in period 8, only a little more than 1% of the total citations by the two domains are between these domains. The intra-domian citations also outnumber the strong inter-domain citations but for a small domain like MRI, for period 8 only by a factor of 5. In the different time periods, 2% to 9% out of 406 possible links in the networks are stronger than would be expected by chance. Thus, their distributions are highly left-skewed over all periods. This indicates that the cross-TD networks are sparsely connected; knowledge flow between *partciular* TDs occurs infrequently despite the fact noted in section 3.1 that intra-domain citations comprise only ~0.3 of the patent to patent citations.

Table 2. Basic statistics of CS values and links in each period

|  | T1 | T2 | T3 | T4 | T5 | T6 | T7 | T8 |
|---|---|---|---|---|---|---|---|---|
| Average | 0.22 | 0.23 | 0.21 | 0.20 | 0.25 | 0.33 | 0.44 | 0.38 |
| Average of CS less than 1 | 0.01 | 0.03 | 0.04 | 0.07 | 0.07 | 0.08 | 0.10 | 0.12 |
| Stadard.dev | 1.97 | 1.18 | 1.18 | 1.01 | 1.22 | 1.63 | 1.59 | 1.33 |
| Max | 33.94 | 13.13 | 14.87 | 16.34 | 15.74 | 23.18 | 17.21 | 13.42 |
|  | EM-FLY | EM-FLY | BAT-FC | BAT-FC | BAT-FC | EM-FLY | BAT-FC | CT-MRI |
| *Number of links whose CS is* | | | | | | | | |
| larger than 1 (strong) | 10 (2%) | 15 (4%) | 15 (4%) | 15 (4%) | 20 (5%) | 28 (7%) | 36 (9%) | 29 (7%) |
| between 0 and 1 (weak) | 15 | 69 | 118 | 166 | 196 | 245 | 255 | 283 |
| Zero | 381 | 322 | 273 | 225 | 190 | 133 | 115 | 94 |

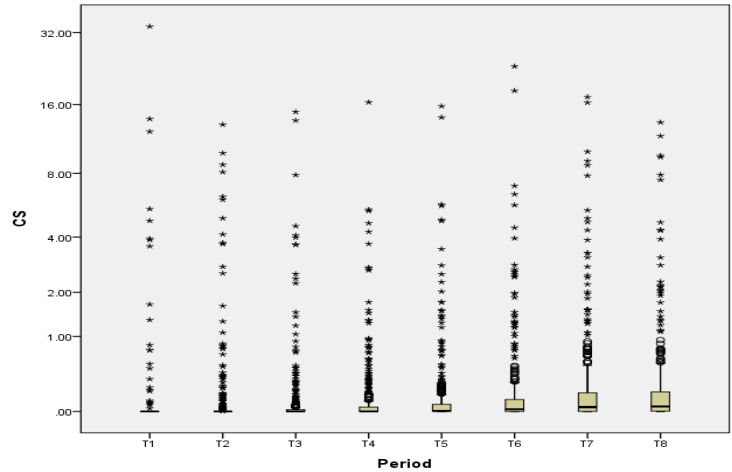

Figure 1. Boxplots of CSs in each period (log scale)

Nonetheless, the networks evolve to become denser over time. There is an increase in strong links but even more increase in weak links. The average value for weak links are small ranging from 0.01 to 0.1 but increase gradually over time. The increasing density arises because backward citations in the patent system have increased over time (Triulzi *et al.*, 2017) and that is a further reason why strong links (above random expectation) is the proper focus for empirical analysis and prediction.

*3.3. Emergence and stability of cross-TD links*

For the 51 strong links, we investigated when they first emerge, how stable they are, and what functional classes they are associated with (Table 3 and 4 gives these results). In terms of emergence, the strong links first equally appear gradually from weak status (21 cases) or abruptly from zero (20 cases). The other 10 strong links are present in the first period and thus the previous state is unknown.

Table 3. Number of new strong links according to emergence, stability, and functional class

|  | T1 | T2 | T3 | T4 | T5 | T6 | T7 | T8 | Total |
|---|---|---|---|---|---|---|---|---|---|
| *Emergence* | | | | | | | | | |
| weak in previous time period | - | 0 | 2 | 3 | 5 | 5 | 6 | 0 | 21 (41%) |
| zero previously | - | 6 | 2 | 1 | 2 | 3 | 3 | 3 | 20 (39%) |
| *Stability* | | | | | | | | | |
| stays until last period | 5 | 2 | 1 | 1 | 4 | 3 | 2 | - | 18 (38%) |
| fluctuates in and out | 5 | 4 | 3 | 3 | 3 | 5 | 7 | - | 30 (63%) |
| *Functional class* | | | | | | | | | |
| within same function | 2 | 0 | 2 | 1 | 3 | 1 | 1 | 1 | 11 (22%) |
| between different operand | 2 | 2 | 0 | 1 | 1 | 2 | 4 | 0 | 12 (24%) |
| between different operation | 4 | 3 | 0 | 0 | 2 | 3 | 2 | 1 | 15 (29%) |
| between different operand and operation | 2 | 1 | 2 | 2 | 1 | 2 | 2 | 1 | 13 (25%) |
| *Total* | 10 | 6 | 4 | 4 | 7 | 8 | 9 | 3 | 51 |

Table 4. List of new strong links according to emergence, stability, and functional class

|  | Spillover within same function | Spillover between different operand | Spillover between different operation | Spillover between different operand and operation |
|---|---|---|---|---|
| T1 | CAM-IC (8)* MIS-OIS (8)* | IC-LED (7) IC-SPP (8)* | BAT-FC (8)* CAM-SIS (3) EM-FLY (8)* IC-SIS (5) | CAP-ET (1) IC-SCD (6) |
| T2 | - | LED-PLG (2)+ CAM-PLG (1)+ | EM-PMM (7)+* PMM-SCD (3)+ LED-SCD (1)+ | AIR-WIND (7)+* |
| T3 | LED-SPP (4) CT-MRI (6)+* | - | - | EM-MIS (5) PMM-MRI (5)+ |
| T4 | EM-WIND (4) | CAM-LED (3)+ | - | EM-ET (1) ET-IL (5)* |
| T5 | IL-LED (4)+* PLG-3D (4)* BAT-CAP (4)* | MIS-PMM (2) | FLY-SCD (1)+ CAP-SCD (2) | CAP-IC (4)* |
| T6 | ET-OT (3)* | IL-3D (2)+ EM-MIL (3)* | EC-SIS (2) PMM-WIND (2)+ BAT-SPP (3)* | BAT-ET (2) FLY-MIL (1)+ |
| T7 | SPP-WIND (1) | CT-3D (1) CAM-IL (1)+ IC-3D (1) SPP-3D (1)+ | EM-SCD (1) CE-FLY (2)* | MRI-SCD (2)* CAP-3D (1)+ |
| T8 | MIL-3D (1)+ | - | FLY-WIND (1)+ | SCD-3D (1)+ |

+Noted links emerge from zero (emergence); numbers in brackets are the counts of the periods when links stay strong and *noted links stay strong until last period (stability)

In terms of stability, the strong links either stay strong until the end or fluctuate as strong, weak, and zero links. 18 among 51 links (38%) stay strong ever since their emergence while 30 links fluctuate, with the last 3 links unclassified since the next state is unknown. The average periods in which links stay strong during the 8 periods is 3.3 periods for all strong links and 5.2 periods for stable links. Even in the case of fluctuating links, they stay strong for an average of 2.2 periods. Over all the time periods, there are five links that always stay strong: EM-FLY, BAT-FC, IC-SPP, MIS-OIS, and CAM-IC. These represent closely associated or partially merged domain pairs for the entire 38 years rather than merging at some point and none of the cases are technologically surprising as being linked. There are four interesting cases where links abruptly emerge and stay stable which will be examined more closely when we test the link prediction methodology: EM-PMM, AIR-WIND, CT-MRI, and IL-LED.

We also investigated functional classes by differentiating links appearing within the same functional class, between different operands, between different operations, or between different operand and operation. For example, both IL and of 3D have transformation as the operation, but the operand of IL is energy and that of 3D is material; thus, IL-3D is spillover between different operands. The portion of the four types that experience strong knowledge flow are similar, although links between different operation capture slightly more. The spillover around the superconductor domain (SCD) occurs frequently between different operations. Integrating with stability, there are many links (7 out of 11) within the same function that are stable. Overall, these results suggest that putting higher weights on functional relationships is not a viable way to improve the predictions and that functional similarity is not useful in predicting knowledge flow.

## 4. Cross-TD network evolution analysis using link prediction

*4.1. Experimental procedure*

For the constructed snapshots of cross-TD networks[3], link prediction metrics are tested to determine their ability to capture the characteristics of the network dynamic evolution. Patent citation-based knowledge flows among TDs for 8 periods form undirected networks with CS weights as documented in the previous section. Link prediction is now conducted in a "back-casting" mode and results compared to the actual results. In this setting, we define a training period as T1, T2,..., and T7 and the testing period as the next period after the training period; if the training network is $G_{T1}$, the testing network is $G_{T2}$. Thus, the link prediction experiments are implemented for 7 periods. Unlike many other studies that predict newly appearing links only, we consider *both new links and recurring links* because the snapshot of cross-TD relationships can fluctuate after first emergence as seen in the previous section.

The overall approach to the experiments are shown in Figure 2. First, the similarity scores for all pairs of TDs are calculated using each of the eight metrics: weighted CommonNeighbors, weighted Jaccard, weighted Adamic-Adar, weighted ResourceAllocation, weighted PreferentialAttachment, Katz, RootedPageRank, and SimRank. The python-based package, 'linkpred' (Guns, 2014) was used for calculation. For global predictors that need to determine parameters ($α, β, γ$), we set several different parameters and examine sensitivity of model performance by trial-and-error. If there is no more change in performance when the parameter increases or decreases, we stop varying the parameter. As a result, $α$ of RootedPageRank and $γ$ of SimRank are set as 0.01, 0.1, 0.5, and 0.9; $β$ of Katz is set as 0.001, 0.1, and 0.5.

Second, the calculated similarity scores are used as weights for a predicted network, which is the basic and most commonly used unsupervised method in link prediction (Murata and Moriyasu, 2007; Newman, 2001). In general link prediction, the aim is to find "existence" of links based on the ranks of similarity values (Zhu and Xia, 2016; Lü and Zhou, 2011). But the objective of the link prediction here is to find "whether the links will have weights larger than 1" that is whether a link greater than randomly expected will exist. A problem arises on the scale of measures: although the actual links have CS values ranging from 0 to ∞, the Jaccard measure ranges from 0 to 1, and PreferentialAttachement ranges for even larger intervals. Thus, the rescaling of similarity scores to be similar to actual CS values is necessary. The rescaling is achieved by normalizing similarity scores (dividing similarity scores by their maximum value) and multiplying by the maximum CS value at the training period. For example, the maximum Adamic-Adar value for $G_{T6}$ is 14.5 and the actual maximum CS of $G_{T6}$ is 23.18; we normalize Adamic-Adar scores for all links by dividing them by 14.5 and then multiply 23.18 so that the predicted link weights are scaled similarly as the actual CSs.

---
[3] See cross-TD networks in appendix 1 (training networks) and 2 (testing networks).

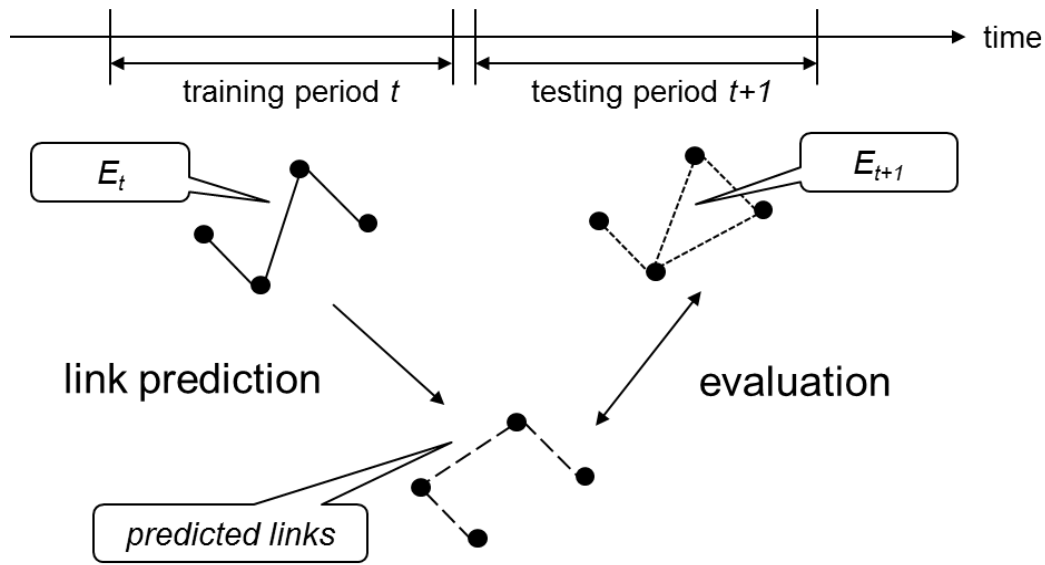

Figure 2. Logical structure of our experiments (modified from Murata and Moriyasu (2007))

Last, the prediction performances are evaluated for each predictor by comparing predicted links and actual links at each testing period, e.g., the predicted link weights calculated from $G_{T6}$, is compared with CS of $G_{T7}$. In particular, as our focus is on the links whose weights are estimated as strong (larger than 1), the confusion matrix (Powers, 2011) is based on whether the CS value is strong (true condition positive) and the predicted link weight is strong (predicted condition positive). Thus, true positives (TPs) are the case when the predicted weight and actual CS of a particular link at a certain testing period are strong and true negatives (TNs) are the opposite case when both are weak or zero. The false positivies (FSs) are the case when a prediction of strong positive link is made for a link that is actually weak or zero whereas the false negatives (FNs) are the case when a prediction that a link is weak or zero is made for a link that is actually strong.

The *accuracy* of the predictors/models measures the ratio of true predictions, both TPs and TNs, over all possible links. The *precision* (positive prediction value) measures whether the links, that a model predicts as strong, are observed to be strong: it is the ratio of the number of TPs divided by predicted condition positive, the total number of links whose weight are strong in the predicted networks. On the other hands, the *recall* (sensitivity) indicates whether a model correctly senses the actual strong links: it is calculated as the number of true positives divided by the total number of links who are actually strong in the testing network. The precision and recall present trade-offs, thus, we find the mean of recall and precision values, which gives us an estimate of how well each predictor represents the actual strong links. The *F-score* of each predictor at each testing period is calculated by taking the harmonic mean of the recall and precision. When |TP|, |TN|, |FP| and |FN| represent the number of true positives, true negatives, false positives and false negatives rates, respectively, the performance measures are formulated as:

$$Accuracy = \frac{|TP|+|TN|}{|TP|+|TN|+|FP|+|FN|} \quad (10)$$

$$\text{Precision} = \frac{true\_positive}{predicted\_condition\_positive} = \frac{|TP|}{|TP|+|FP|} \quad (11)$$

$$\text{Recall} = \frac{true\_positive}{true\_condition\_positive} = \frac{|TP|}{|TP|+|FN|} \quad (12)$$

$$F-score = \frac{2 \times \text{Precision} \times \text{Recall}}{\text{Precision}+\text{Recall}} \quad (13)$$

*4.2. Testing results for predictions*

The performance of all of the predictors[4] for all periods are illustrated in Figure 3. The accuracy of the Katz models (figure 3a) are close to 1 with particularly small deviation at $\beta$=0.001 while that of RootedPageRank, PreferentialAttachement, CommonNeighbors, and Adamic-Adar are also >0.8. The precision of the Katz models (figure 3b) are close to 0.8 while that of RootedPageRank, PreferentialAttachment, CommonNeighbors, and Adamic-Adar are <0.4. But the recall (sensitivity) (figure 3c) shows somewhat different results: the recall of RootedPageRank models are >0.8 especially at $\alpha$ = 0.9 and 0.5; that of the SimRank models at a large $\gamma$ = 0.9 and 0.5 are also close to 0.8; and that of ResouceAllocation is 0.7 with large deviation whereas that of the Katz at $\beta$=0.001 is close to 0.7 with small deviation. Consequently, for the key overall F-score (figure 3d), the Katz ($\beta$=0.001) model is the best performer at every period, vastly superior to other predictors. The Katz ($\beta$=0.01) model has the same performance at T2, T4, and T6 but lower F-score in other periods.

---

[4] For detailed values, see Appendix 4

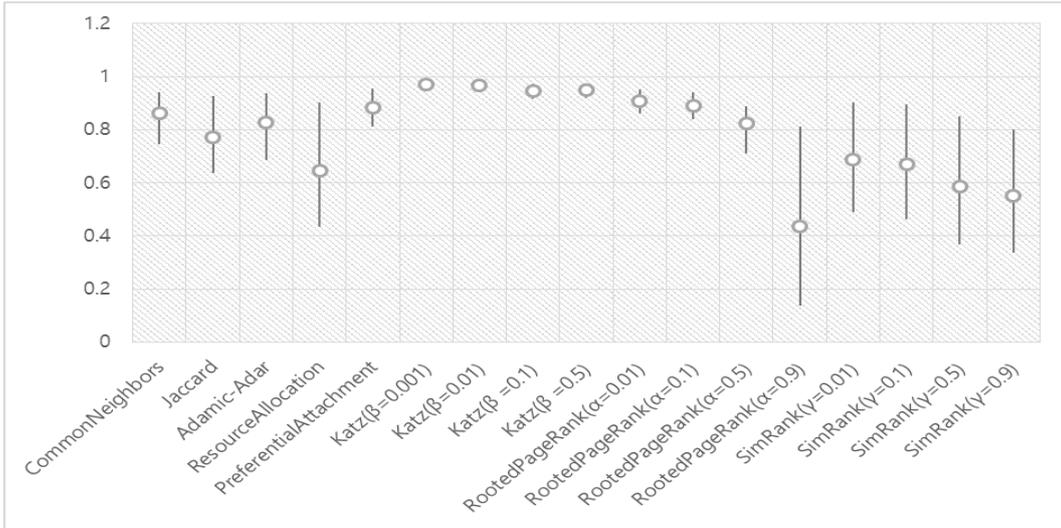

(a) Accuracy

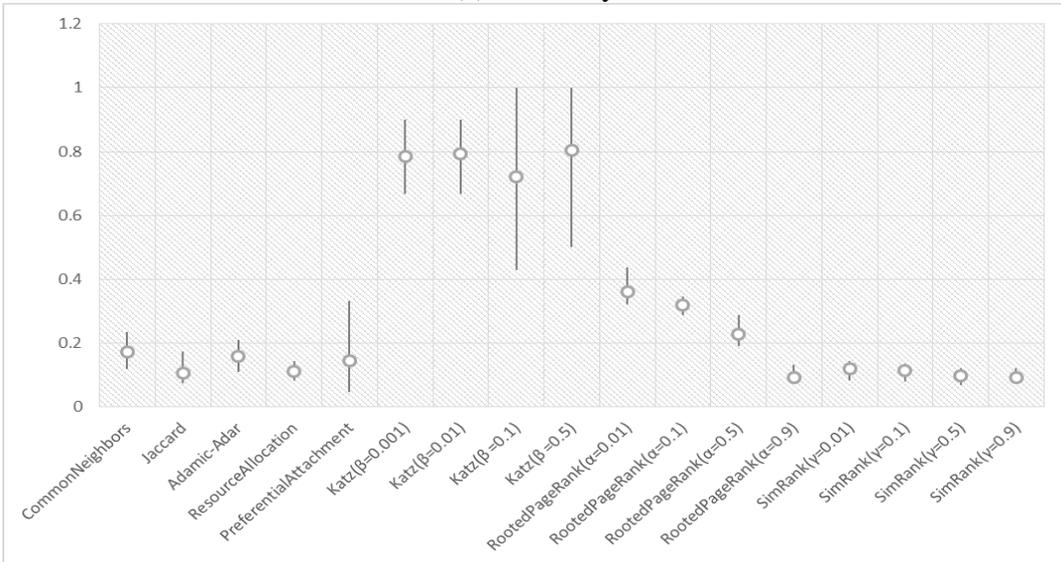

(b) Precision

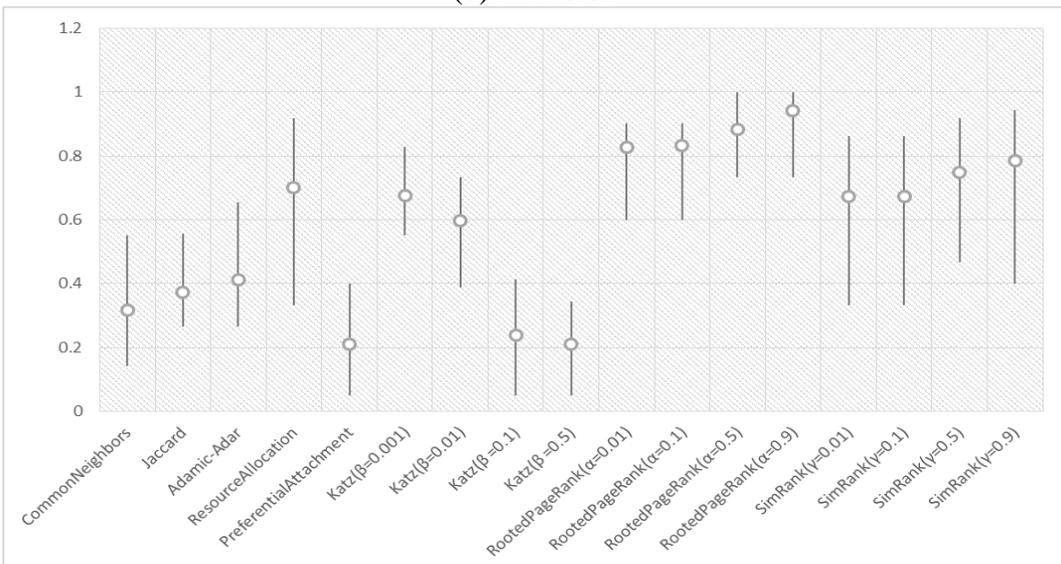

(c) Recall

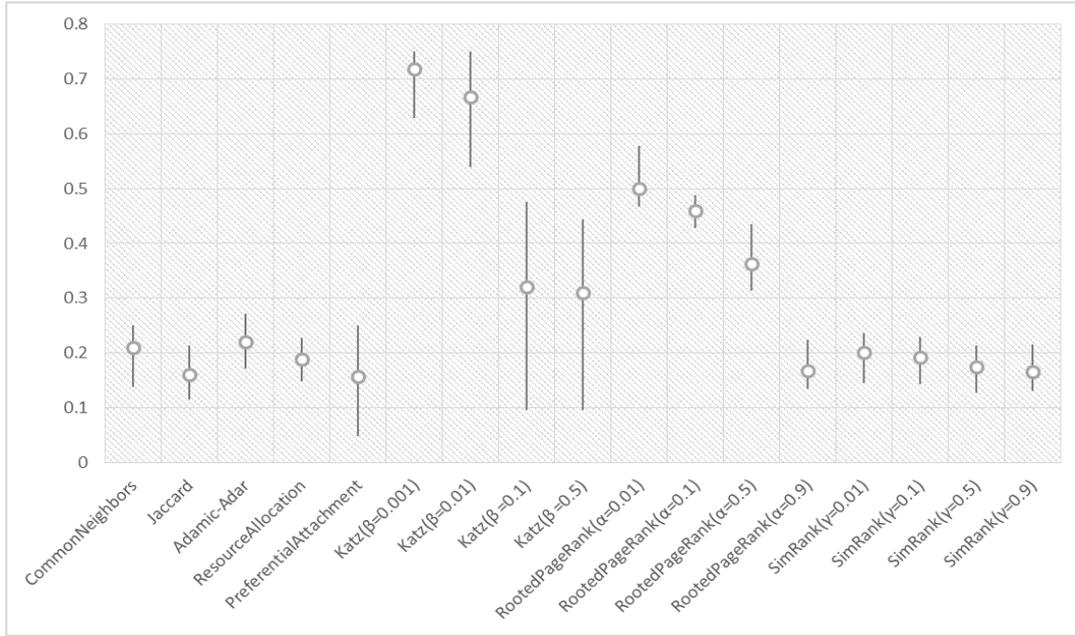

(d) F-score

Figure 3. Link prediction performance: (a) accuracy; (b) precision; (c) recall; and (d) F-score

Although recall is not so high, the Katz model turns out as the best performer due to the high precision it achieves. If the Katz predicts that a link is strong, the prediction is almost always correct. But the Katz predictions can overlook or not predict some actual strong links. In contrast, the RootedPageRank model has better recall (~0.8) than Katz (~0.7) but has substantially poorer precision (~0.4) than Katz (~0.8). Due to good recall, the RootedPageRank senses or detects the actual strong links well. However, as the poor precision means that many case of predictions are false alarms, the strong links that the model predicts are often actually not strong. Similar results of large recall and small precision are also found in SimRank and ResouceAllocation. The Katz metric predicts only a small number of both strong and weak links whereas the others predicted large number of both types[5] although their contribution to TPs are different.

As the Katz model yields a small number of positive predictions, its TPs are also relatively small. We found that most TPs of the Katz model are also predicted by RootedPageRank. Going beyond the prediction of strong or not, Figure 4[6] illustrates the actual and predicted link weights for four interesting links that abruptly emerge and stay stable. The CT-MRI (figure 4a) and AIR-WIND (figure 4b) show excellent performance by the Katz model at all periods, except for a FN at the start of increases in T3. The RootedPageRank also shows decent performance despite overly large values, but other metrics are poor. In a case when peak patterns have stability as in the IL-LED (figure 4c), both predictors also perform well. In summary the RootedPageRank tends to overestimate the link weights and have bigger increases than the Katz, with the exception of the EM-PMM case (figure 4d): the RootedPageRank is slightly passive in following the actual patterns, even though both models are successful at all periods, except for T2.

---

[5] For detailed number of predicted links, see Appendix 3

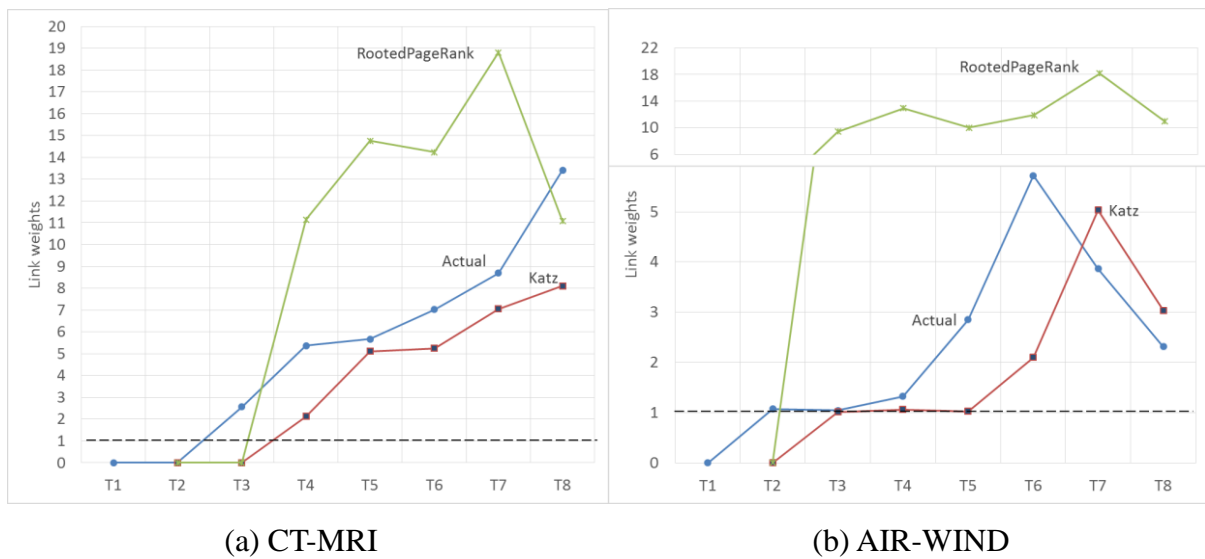

(a) CT-MRI  (b) AIR-WIND

---

[6] Note that the global models in figures 4-6 are the best performing parameters: Katz ($\beta=0.001$) and RootedPageRank ($\alpha=0.01$)

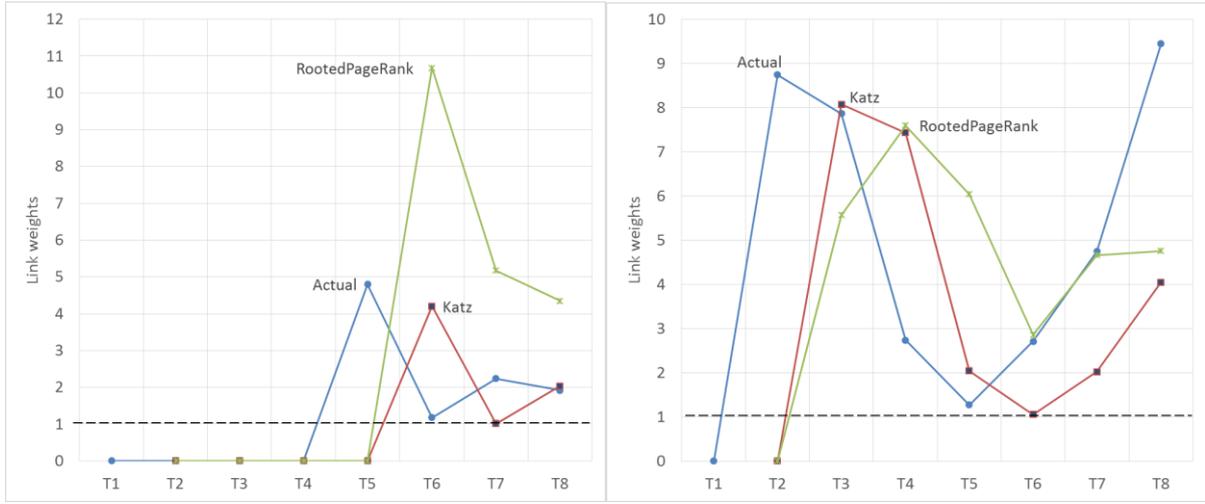

(c) IL-LED  (d) EM-PMM

Figure 4. Some cases of links abruptly emerge and stay stable: successful in both Katz and RootedPageRank

Figure 5 shows other cases of successful predictions of the Katz model only. In the LED-PLG case (figure 5a), the model successfully predicts being stable at T3 and the decrease at T5, even though unsuccessful in detecting the decrease in T4 (FP). RootedPageRank also predicts a decrease but largely misses the degree of change. The CAP-ET (figure 5b) and CAM-SIS (figure 5c) start as strong links but decrease to weak links in early periods (T2 and T3, repectively). In these cases, the Katz estimates rapidly decrease and stay zero in remaining periods: although it misses the first decrease in T3 and T4 (FP), there are no FNs. But the RootedPageRank model stays strong in every period thus there are many FPs. When comparing Katz with CommonNeighbors in Figure 5a and 5b, they show compeletely different predictions. Although the Katz model with a very small $β$ might be intuitively expected to yield a measurement close to CommonNeighbors (because the long paths then contribute very little), in fact Katz with small beta emphasizes the direct link whereas CommonNeighbors does not instead emphasizing the links between neighboring nodes.

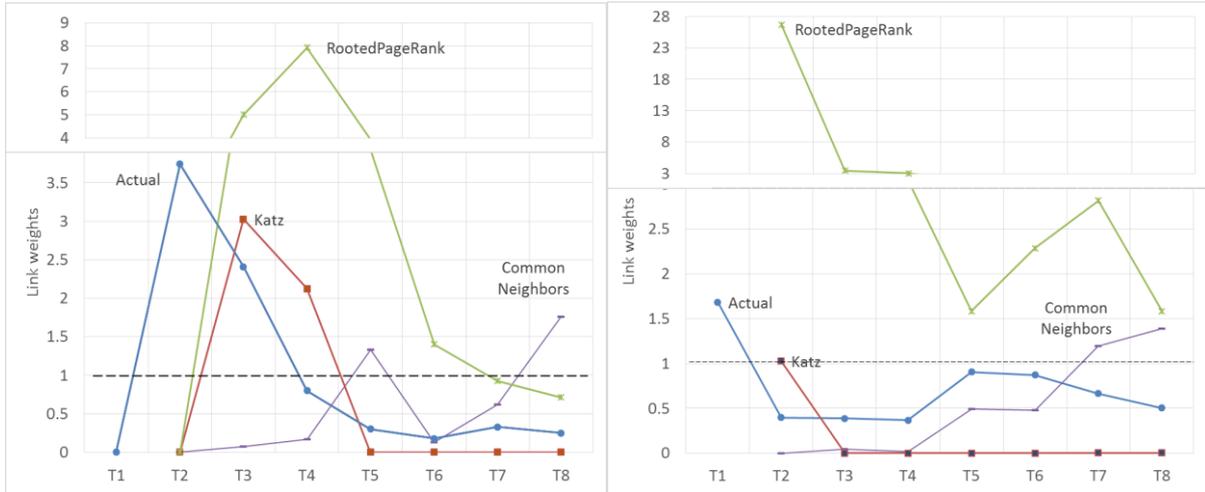

(a) LED-PLG    (b) CAP-ET

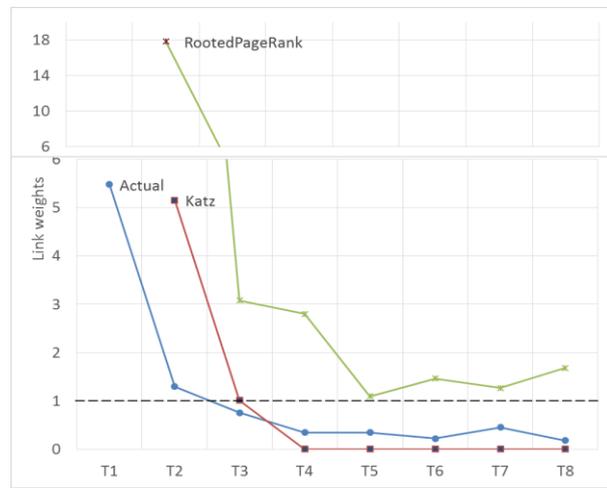

(c) CAM-SIS

Figure 5. Some cases of successful predictions of the Katz model

On the other hand, Figure 6 presents unsuccessful predictions of the Katz model. In the PMM-MRI case (figure 6a), although both the Katz model and RootedPageRank fail to predict the decrease at T4, RootedPageRank predicts successfully after that while the Katz model has a TP at only T7 and T8. In the CE-FLY case (figure 6b), the Katz model continues to show negative prediction and has a TP only at T8 whereas the RootedPageRank begins to increase at T5, and shows TPs at T6 to T8 and T7 to T8 respectively. Similarly, in the EC-SIS case (figure 6c), the TPs of Katz exist only at T7 whereas that of the RootedPageRank is at T6 and T7. But in two cases of CE-FLY and EC-SIS, RootedPageRank has many FPs.

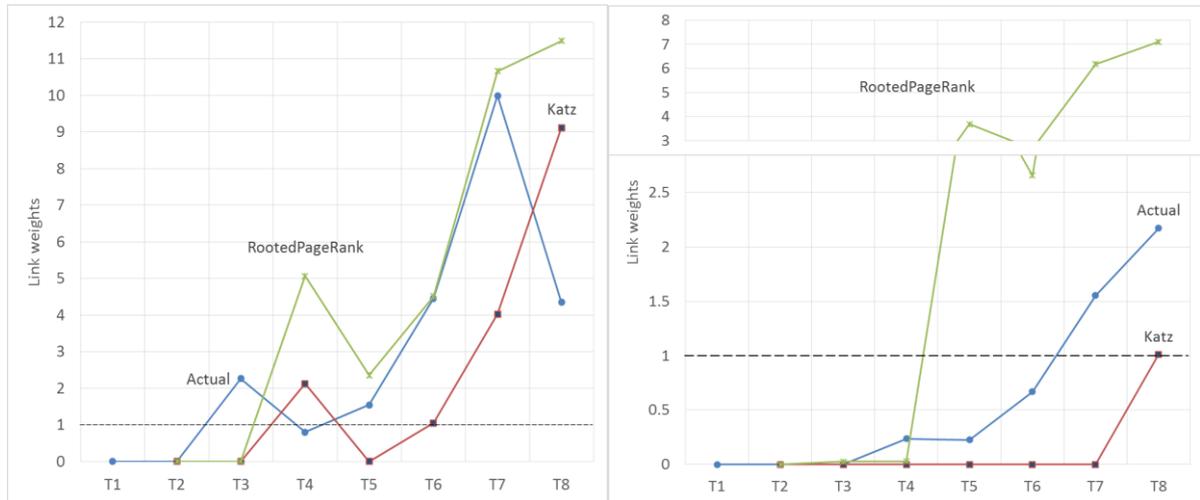

(a) PMM-MRI    (b) CE-FLY

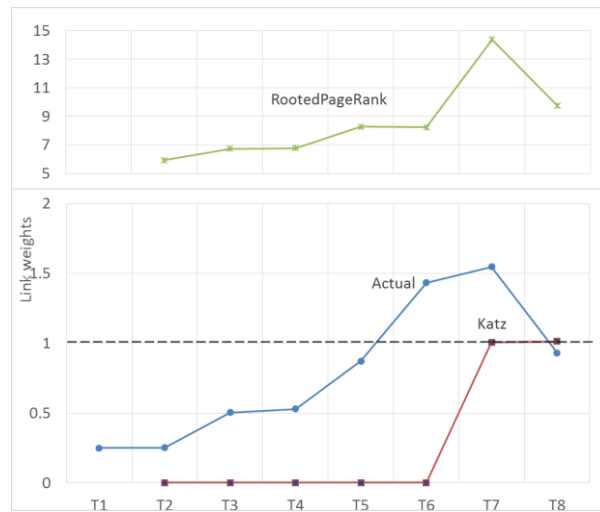

(c) EC-SIS

Figure 6. Some cases of unsuccessful predictions of the Katz

Whether successful or not, the Katz model tends to predict the increase or decrease of link weights in a conservative and cautious manner while the RootedPageRank models predict in an excessively positive way. The Katz model is highly dependent on the last CS that is larger than 1; it follows the training data trends with one period of time lag. It is successful in predicting sustaining, stable links; but weak in predicting new emergence of strong links, which emerge abruptly (from zero) or incrementally (from weak), because the model rarely considers weak links.

## 5. Discussion

This paper empirically analyzed the dynamic pattern and evolution of technological knowledge flows among TDs. A key empirical finding is that the domains we studied interact with only a small minority of other specific domains (2% in the initial period and 9% at the highest in period 7). Moreover, the fraction of citations going to *any other specific* domain is less than a few percent of the total patent citations by a domain, but on average 70% of the patent-citations are to patents not in the domain of the patent. Thus, empirically it is clear that cross-domain knowledge flow is important overall (~70% of knowledge input on average) even though it is restricted somewhat to specific domain pairs since only 2-9% of possible links are strong in our results. A further implication is that the average domain must interact with *multiple* other domains in order to achieve the 70% inter-domain knowledge flow since typical strong inter-domain knowledge flows for a specific linkage only account for a few percent of the total knowledge flow to or from a domain..

In regard to the empirical time dependence of the cross-domain linkages, most that are above expectation become relatively steady after emerging from either zero or weak interactions. Some show slow decreases in intensity, some show slow increases and others are classified as steady (see Table 5). The average of 83% condition positives (strong links of testing networks at each period) follow decreasing, increasing or steady patterns. Table 5 also shows that among all condition positives, 68% of the strong links at T were also strong at T-1. This finding suggests that interaction among domains is not "one-time knowledge flow" but mostly continual idea transfer. We also find that knowledge flow is heterogenous and is almost equally likely to occur across operands (from energy technologies to information technologies) and across operations (from storage to transformation) as is it to occur within a given functional classification (for example energy storage). Overall, these empricial results suggest that the limits to transfer of ideas between domains (small numbers of strong interactions) are not functionally dictated.

Table 5. Interpretation (pattern proportion)

| Condition positives | T2 | T3 | T4 | T5 | T6 | T7 | T8 | Total |
|---|---|---|---|---|---|---|---|---|
| Total | 15 | 15 | 15 | 20 | 28 | 36 | 29 | 158 |
|   Decreasing | 7 | 6 | 4 | 5 | 4 | 5 | 4 | 35 |
|   Increasing | 0 | 2 | 3 | 5 | 8 | 9 | 9 | 36 |
|   Steady | 5 | 7 | 8 | 7 | 10 | 10 | 8 | 55 |
|   Peak | 3 | 0 | 0 | 3 | 6 | 12 | 8 | 32 |
| Proportion of decreasing, increasing, steady patterns | 80% | 100% | 100% | 85% | 79% | 67% | 72% | 83% |
| Links that are also strong at T-1 | 9 (60%) | 11 (73%) | 11 (73%) | 11 (55%) | 18 (64%) | 23 (64%) | 24 (83%) | 107 (68%) |

An important distinction of the work reported here is that we pursue prediction of links in the next time period based upon observations in the preceding time period. This has practical implications that we will discuss further below. At this point, we note that our testing of various models for such prediction suggests additional implications for understanding knowledge flow among differing technological domains. An important finding of this type is that triadic closure is unimportant as a mechanism for forming linkages among technological domains despite its widespread utility for understanding the dynamic evolution of many complex networks (Bianconi *et al.*, 2014; Kleinberg, 2008; Liben-Nowell and Kleinberg, 2007). This is shown by the non-existent prediction power of the CommonNeighbor and other local topology prediction models. The fact that two technologies have significant interaction with a specific third domain not at all predicts that these two technologies will interact in the future. For example, the camera and LED lighting domains were not linked but had a strong common neighbor integrated circuit (IC) domain at T1 (the CS of CAM-IC was 13.89 and that of IC-LED was 4.81). Not surprisingly, the CAM-LED link is predicted as strong by most local predictors, but the actual CS at T2 was zero. The links between semiconductor technology (IC) and both LEDs and Camera1 sensitivity are aligned with technical facts about these domains, but there are no such technical reasons for knowledge flow directly between LEDs and Camera development. Among many other similar examples, we note flywheel and wind turbines (FLY-WIND) at T5 where having both magnetic information storage (MIS) and electric motors (EM) as common neighbors leads to a predicted link; but since these common neighbors do not signify a knowledge flow between flywheels and wind turbines, the predicted link does not occur. If we contrast this result with the very successful Katz prediction at low Beta, we are led to the conclusion that the Katz predictor reliance on direct links to predict further interaction overcomes the weakness of triadic closure prediction, mainly by assuring an ability to interact directly before the prediction of stronger interaction.

Similar to the failure of triadic closure in being an effective predictive mechanism, it appears that the "rich get richer" mechanism associated with power laws and preferential attachment (Barabási *et al.*, 2002; Watts and Strogatz, 1998) is not effective in predicting future links. The F-score and other measures of the effectiveness of the power law metrics is weak (figures 3) and suggests that preferential attachment is not the mechanism by which domains are linked to one another. Thus, domains that have a relatively high number of interactions at one time period are not more likely to form new linkages in the next time period. Interestingly, this suggests that the conceptually appealing idea of general purpose technologies as knowedge sources (Moser *et al.*, 2015; Jovanovic and Rousseau, 2005) is not supported: general purpose technologies would be high degree nodes in a cross-domain network, and if general purpose technologies are thus a dominant mechanism for the spillover of ideas, preferential attachment would be expected to show good performance. The results do not demonstrate such an effect. This result does not suggest that general purpose technologies are economically unimportant but instead that extending the concept to cover knowledge spillovers is not supported by our results. There is good evidence (Brynjolfsson and McAfee, 2014) that the integrated circuits domain has been very economically significant during the last 40 years; nonetheless, the inclusion of IC in the 29 domains we investigated did not result in importance of highly interactive domains in forging new knowledge-flow links over time.

The most successful predictors are global metrics (Katz and RootedPageRank) based upon the overall network that do not assume near neighbor interactions as important. This finding suggests that paths of strong cumulative development such as first suggested by Dosi are a viable concept. Moreover, the success of the Katz model is not just relatively strong because of the failiure of triadic closure and preferential attachment but absolutely strong enough to be practically useful: The Katz model successfully predicted when a link stably stays strong (as shown in Figure 4) and predicted better than RootedPageRank when a link changes to and stays stable at weak or zero (as shown in Figure 5). This absolute strength is consistent with the steady patterns exhibited by most strong interactions among the domains and with the fact that strong links continue strong in next period (as shown in Table 5). Thus, the Katz model gives us a quantitative, predictive interpretation of why the strong linkages are relatively steady with time. Qualitatively, the success of the model indicates that most of the inter-domain knowledge flow is continuation of past flows starting from when domains are first formed, and also suggests that trajectories utilize technologically diverse knowledge.

## 6. Conclusion

This study investigated the dynamic evolution of knowledge flow across technological domains. Filling the gaps from the previous research, the study used technological domain units to capture knowledge flows among technological artifacts, and the link prediction method to help characterize the topological change and evolutionary behaviors in dynamic technology networks. As an early attempt that applies link prediction to patent citation-based networks, this study offers the basis for predicting the emergence and continuance of future technological knowledge flows.

This research particularly has implications for the dynamic capabilities of firms (Teece, 2007; Teece, Pisano, and Shuen, 1997). Dynamic capabilities allow significant change to occur within the firm so that it can adapt more effectively than other firms to a change in the environment (Teece, 2007). Given the importance of cross-domain knowledge spillovers to creation of technological variety (Schoenmakers and Duysters, 2010), a potentially important aspect of firm dynamic capabilities is the anticipation and integration of such spillovers to domains that are critical to the firm. Our findings show that the existence of such spillovers is predictable to a significant extent through link prediction using the Katz model. Thus, our results suggest that a key strategic asset of a firm can be at least somewhat planned for.

A first task for any firm striving to be more mindful of important sources of knowledge spillovers is to first analyze which technological domains are important in the products and services they produce. A systematic procedure for identifying more details about such technologies is available (Benson and Magee, 2015, 2016) and can lead to identification of patents with all of their attendant information, such as citations by these patents, which is the key step in identifying the important knowledge spillover sources. Having the capability to handle such spillovers will require some attention to expertise acquisition (employees or outsiders) and/or collaboration in the technological domains of importance, but also some attention to developments in "non-core" areas.

Our results indicate that the sources of the knowledge spillover are relatively stable, making identification and awareness maintenance more doable for the firms who take it on.

Such stability in a key aspect of long-term change was pointed out in an important paper by Helfat and Winter (2011), which showed that dynamic and operational capabilities had at best a fuzzy boundary. In the case discussed here, the spillover domains are mostly fixed over time, with only the formation of new domains unknown.

In addition to further research on the dynamics of the formation of new domains, other aspects of the research reported here that might be improved in the future are worth noting. First, while the patent coverage in our dataset is extensive (502,444 patents), it is not the total US patent set for the years studied (4,666,574 patents) and thus define samples of domains rather than a total set. This is important since the focus of the work is cross-domain knowledge flow; however, the set of domains tested here is numerous and broad enough that our results (network metrics that are good at predicting relatively stable linkages work best) appear robust. While we tested a wide range of models, first difference time series have recently been demonstrated for link prediction (for example using ARIMA) and future research could explore this.

**Acknowledgments**

The authors acknowledge support from the SUTD-MIT International Design Center (IDC). We also thank Chris Benson and Subarna Basnet for sharing the patent sets for the 29 technology domains used in this study.

**References**

Acemoglu D, Akcigit U, Kerr WR. 2016. Innovation network. *Proceedings of the National Academy of Sciences* **113**(41): 11483–11488.
Adamic LA, Adar E. 2003. Friends and neighbors on the Web. *Social Networks* **25**(3): 211–230.
Alcácer J, Gittelman M. 2006. Patent Citations as a Measure of Knowledge Flows: The Influence of Examiner Citations. *Review of Economics and Statistics*. The MIT Press **88**(4): 774–779.

Barabási A. *et al.* 2002. Evolution of the social network of scientific collaborations. *Physica A: Statistical Mechanics and its Applications* **311**(3): 590–614.
Basnet S. 2015. *Modeling Technical Performance Change Using Design Fundamentals*. Massachusetts Institute of Technology.
Basnet S, Magee CL. 2016. Modeling of technological performance trends using design theory. *Design Science* **2**(e8).
Benson CL, Magee CL. 2013. A hybrid keyword and patent class methodology for selecting relevant sets of patents for a technological field. *Scientometrics* **96**(1): 69–82.
Benson CL, Magee CL. 2015. Technology structural implications from the extension of a patent search method. *Scientometrics* **102**(3): 1965–1985.
Benson CL, Magee CL. 2016. Using Enhanced Patent Data for Future-Oriented Technology Analysis. In *Anticipating Future Innovation Pathways Through Large Data Analysis*, Tugrul U. Daim, Denise Chiavetta, Alan L. Porter, Ozcan Saritas (eds). Springer International Publishing: 119–131.
Bianconi G, Darst R, Iacovacci J, Fortunato S. 2014. Triadic closure as a basic generating mechanism of communities in complex networks. *Physical Review E* **90**(4): 42806.
Brin S, Page L, Brin S, Page L. 1998. The anatomy of a large-scale hypertextual Web search engine. *Computer Networks and ISDN Systems*. Elsevier Science Publishers B. V. **30**(1–7): 107–117.
Brynjolfsson E, McAfee A. 2014. *The Second Machine Age: Work, Progress, and Prosperity in a Time of Brilliant Technologies*. W. W. Norton & Company: New York.
Caviggioli F. 2016. Technology fusion: Identification and analysis of the drivers of technology convergence using patent data. *Technovation* **55–56**: 22–32.
Chang S-B, Lai K-K, Chang S-M. 2009. Exploring technology diffusion and classification of business methods: Using the patent citation network. *Technological Forecasting and Social Change* **76**(1): 107–117.
Chen Z, Guan J. 2016. The core-peripheral structure of international knowledge flows: evidence from patent citation data. *R&D Management* **46**(1): 62–79.
Cho T-S, Shih H-Y. 2011. Patent citation network analysis of core and emerging technologies in Taiwan: 1997–2008. *Scientometrics* **89**(3): 795–811.
Cho Y, Kim E, Kim W. 2015. Strategy transformation under technological convergence: evidence from the printed electronics industry. *International Journal of Technology* **67**(2/3/4): 106–131.
Choi C, Park Y. 2009. Monitoring the organic structure of technology based on the patent development paths. *Technological Forecasting and Social Change* **76**(6): 754–768.
Choi J, Hwang Y-S. 2014. Patent keyword network analysis for improving technology development efficiency. *Technological Forecasting and Social Change* **83**: 170–182.
Clauset A, Moore C, Newman MEJ. 2008. Hierarchical structure and the prediction of missing links in networks. *Nature* **453**(7191): 98–101.
Clough JR, Gollings J, Loach T V., Evans TS. 2015. Transitive reduction of citation networks. *Journal of Complex Networks* **3**(2): 189–203.
Dasgupta K, Singh R, Viswanathan B. 2008. Social ties and their relevance to churn in mobile telecom networks. In *Proceedings of the 11th international conference on Extending database technology: Advances in database technology*. Nantes, France.
Dosi G. 1982. Technological paradigms and technological trajectories. *Research Policy* **11**(3): 147–162.
Duguet E, MacGarvie M. 2005. How well do patent citations measure flows of technology? Evidence from French innovation surveys. *Economics of Innovation and New Technology*. Routledge **14**(5): 375–393.

Érdi P *et al.* 2013. Prediction of emerging technologies based on analysis of the US patent citation network. *Scientometrics*. Springer Netherlands **95**(1): 225–242.

Esslimani I, Brun A, Boyer A. 2011. Densifying a behavioral recommender system by social networks link prediction methods. *Social Network Analysis and Mining*. Springer Vienna **1**(3): 159–172.

Funk RJ, Owen-Smith J. 2016. A Dynamic Network Measure of Technological Change. *Management Science* (March): mnsc.2015.2366.

Gentner D, Markman A. 1997. Structure mapping in analogy and similarity. *American Psychologist* **52**(1): 45–56.

Guimerà R, Sales-Pardo M. 2009. Missing and spurious interactions and the reconstruction of complex networks. *Proceedings of the National Academy of Sciences of the United States of America*. National Academy of Sciences **106**(52): 22073–8.

Guns R. 2014. Link Prediction. In *Measuring Scholarly Impact - Methods and Practice*, Ding Y, Rousseau R, Wolfram D (eds). Springer International Publishing: Cham: 35–56.

Han Y, Park Y. 2006. Patent network analysis of inter-industrial knowledge flows: The case of Korea between traditional and emerging industries. *World Patent Information* **28**(3): 235–247.

Helfat CE, Winter SG. 2011. Untangling dynamic and operational capabilities: Strategy for the (N)ever-changing world. *Strategic Management Journal* **32**(11): 1243–1250.

Huang Z, Zeng D. 2006. A link prediction approach to anomalous email detection. In *Systems, Man and Cybernetics, 2006. SMC'06. IEEE International Conference on*: 1131–1136.

Hung S-W, Wang A-P. 2010. Examining the small world phenomenon in the patent citation network: a case study of the radio frequency identification (RFID) network. *Scientometrics*. Springer Netherlands **82**(1): 121–134.

Itakura KY, Clarke CLA, Geva S, Trotman A, Huang WC. 2011. Topical and Structural Linkage in Wikipedia. In *Advances in Information Retrieval. ECIR 2011. Lecture Notes in Computer Science, vol 6611*, Clough P et al. (eds). Springer: Berlin, Heidelberg: 460–465.

Jaffe A, Trajtenberg M, Fogarty M. 2000. Knowledge Spillovers and Patent Citations: Evidence from a Survey of Inventors. *American Economic Review* **90**(2): 215–218.

Jaffe A, Trajtenberg M, Henderson R. 1993. Geographic localization of knowledge spillovers as evidenced by patent citations. *the Quarterly journal of Economics* **108**(3): 577–598.

Jeh G, Widom J. 2002. SimRank: a measure of structural-context similarity. In *Proceedings of the eighth ACM SIGKDD international conference on Knowledge discovery and data mining - KDD '02*. ACM Press: New York, USA: 538–543.

Jovanovic B, Rousseau PL. 2005. General Purpose Technologies. In *Handbook of Economic Growth*, Aghion P, Durlauf SN (eds), 1B: 1181–1224.

Karki M. 1997. Patent citation analysis: A policy analysis tool. *World Patent Information*.

Katz L. 1953. A new status index derived from sociometric analysis. *Psychometrika*. Springer-Verlag **18**(1): 39–43.

Kim D, Cerigo DB, Jeong H, Youn H. 2016. Technological novelty profile and invention's future impact. *EPJ Data Science* **5**(8): 1–15.

Kim E, Cho Y, Kim W. 2014. Dynamic patterns of technological convergence in printed electronics technologies: Patent citation network. *Scientometrics* **98**(2): 975–998.

Kleinberg J. 2008. The convergence of social and technological networks. *Communications of the ACM* **51**(11): 66–72.

Ko N, Yoon J, Seo W. 2014. Analyzing interdisciplinarity of technology fusion using knowledge flows of patents. *Expert Systems with Applications* **41**(4): 1955–1963.

Koh H, Magee CL. 2006. A functional approach for studying technological progress:

Application to information technology. *Technological Forecasting and Social Change* **73**(9): 1061–1083.

Lee S, Kim M-S. 2010. Inter-technology networks to support innovation strategy: An analysis of Korea's new growth engines. *Innovation* **12**(1): 88–104.

Lee WJ, Lee WK, Sohn SY. 2016. Patent Network Analysis and Quadratic Assignment Procedures to Identify the Convergence of Robot Technologies. *Plos One*, Gao Z-K (ed). Public Library of Science **11**(10): e0165091.

Li J, Zhang L, Meng F, Li F. 2014. Recommendation Algorithm based on Link Prediction and Domain Knowledge in Retail Transactions. *Procedia Computer Science* **31**: 875–881.

Liben-Nowell D, Kleinberg J. 2007. The link-prediction problem for social networks. *Journal of the American Society for Information Science and Technology* **58**(7): 1019–1031.

Lü L, Zhou T. 2011. Link prediction in complex networks: A survey. *Physica A: Statistical Mechanics and its Applications* **390**(6): 1150–1170.

Luan C, Liu Z, Wang X. 2013. Divergence and convergence: Technology-relatedness evolution in solar energy industry. *Scientometrics* **97**(2): 461–475.

Magee CL, Basnet S, Funk JL, Benson CL. 2016. Quantitative empirical trends in technical performance. *Technological Forecasting and Social Change*. The Authors **104**: 237–246.

Martinelli A, Nomaler Ö. 2014. Measuring knowledge persistence: a genetic approach to patent citation networks. *Journal of Evolutionary Economics*. Springer Berlin Heidelberg **24**(3): 623–652.

Mina A, Ramlogan R, Tampubolon G, Metcalfe JS. 2007. Mapping evolutionary trajectories: Applications to the growth and transformation of medical knowledge. *Research Policy* **36**(5): 789–806.

Moser P, Nicholas T, Nicholas TOM. 2015. Was Electricity a General Purpose Technology? Patent Citations Evidence from Historical. *The American Economic Review, Papers and Proceedings* **94**(2): 388–394.

Murata T, Moriyasu S. 2007. Link Prediction of Social Networks Based on Weighted Proximity Measures. In *IEEE/WIC/ACM International Conference on Web Intelligence (WI'07)*. IEEE: 85–88.

Newman MEJ. 2001. Clustering and preferential attachment in growing networks. *Physical Review E*. American Physical Society **64**(2): 25102.

No HJ, Park Y. 2010. Trajectory patterns of technology fusion: Trend analysis and taxonomical grouping in nanobiotechnology. *Technological Forecasting and Social Change*. Elsevier Inc. **77**(1): 63–75.

Nomaler Ö, Verspagen B. 2016. River deep, mountain high: Of long-run knowledge trajectories within and between innovation clusters. *Journal of Economic Geography* **16**(6): 1259–1278.

Park H, Magee CL. 2017. Tracing Technological Development Trajectories: A Genetic Knowledge Persistence-Based Main Path Approach. *PLoS ONE*, Gao Z-K (ed). Harvard University Press **12**(1): e0170895.

Park J, Lee H, Park Y. 2009. Disembodied knowledge flows among industrial clusters: A patent analysis of the Korean manufacturing sector. *Technology in Society*. Elsevier Ltd **31**(1): 73–84.

Perez-Cervantes E, Mena-Chalco JP, Oliveira MCF De, Cesar R. 2013. Using Link Prediction to Estimate the Collaborative Influence of Researchers. In *2013 IEEE 9th International Conference on e-Science*. IEEE: 293–300.

Powers DMW. 2011. Evaluation: From Precision, Recall, and F-measure to ROC, Informedness, Markedness & Correlation. *Journal of Machine Learning Technologies*

**2**(1): 37–63.

Price D de S. 1976. A general theory of bibliometric and other cumulative advantage processes. *Journal of the American society for Information* **27**(5): 292–306.

Ruttan VW. 1959. Usher and Schumpeter on Invention, Innovation, and Technological Change. *The Quarterly Journal of Economics* **73**(4): 596–606.

Salton G, McGill M. 1983. *Introduction to modern information retrieval*. McGraw-Hill: New York.

Schoenmakers W, Duysters G. 2010. The technological origins of radical inventions. *Research Policy* **39**: 1051–1059.

Shih H-Y, Chang T-LS. 2009. International diffusion of embodied and disembodied technology: A network analysis approach. *Technological Forecasting and Social Change* **76**(6): 821–834.

Shin J, Park Y. 2010. Evolutionary optimization of a technological knowledge network. *Technovation* **30**(11–12): 612–626.

Teece D. 2007. Explicating dynamic capabilities: the nature and microfoundations of (sustainable) enterprise performance. *Strategic management journal* **28**(13): 1319–1350.

Teece DJ, Pisano G, Shuen A. 1997. Dynamic Capabilities and Strategic Management. *Strategic Management Journal*. Wiley **18**(7): 509–533.

Trajtenberg M. 1990. A Penny for Your Quotes: Patent Citations and the Value of Innovations. *The RAND Journal of Economics* **21**(1): 172.

Travers J, Milgram S. 1967. The small world problem. *Phychology Today* **1**(1): 61–67.

Triulzi G, Alstott J, Magee CL. 2017, June 15. Predicting Technology Performance Improvement Rates by Mining Patent Data. *SSRN*. Available at: https://ssrn.com/abstract=2987588.

Usher AP. 1954. *A history of mechanical inventions*. Harvard University Press: Cambridge, MA.

Valverde-Rebaza J, de Andrade Lopes A. 2013. Exploiting behaviors of communities of twitter users for link prediction. *Social Network Analysis and Mining*. Springer Vienna **3**(4): 1063–1074.

Verspagen B. 2007. Mapping technological trajectories as patent citation networks: a study on the history of fuel cell research. *Advances in Complex Systems*. World Scientific Publishing Company **10**(1): 93–115.

Verspagen B, De Loo I. 1999. Technology Spillovers between Sectors and over Time. *Technological Forecasting and Social Change* **60**(3): 215–235.

Ter Wal A. 2014. The dynamics of the inventor network in German biotechnology: geographic proximity versus triadic closure. *Journal of Economic Geography* **14**(3): 589–620.

Watts D, Strogatz S. 1998. Collective dynamics of 'small-world'networks. *Nature* **393**(6684): 440–442.

Weisberg RW. 2006. *Creativity: understanding innovation in problem solving, science, invention, and the arts*. John Wiley & Sons.

Zhou T, Lu L, Zhang Y-C. 2009. Predicting Missing Links via Local Information. *European Physical Journal B* **71**(4): 623–630.

Zhu B, Xia Y. 2016. Link Prediction in Weighted Networks: A Weighted Mutual Information Model. *PLoS ONE*, Sendiña-Nadal I (ed). Public Library of Science **11**(2): e0148265.

## Appendix 1. Cross-TD networks (training networks)

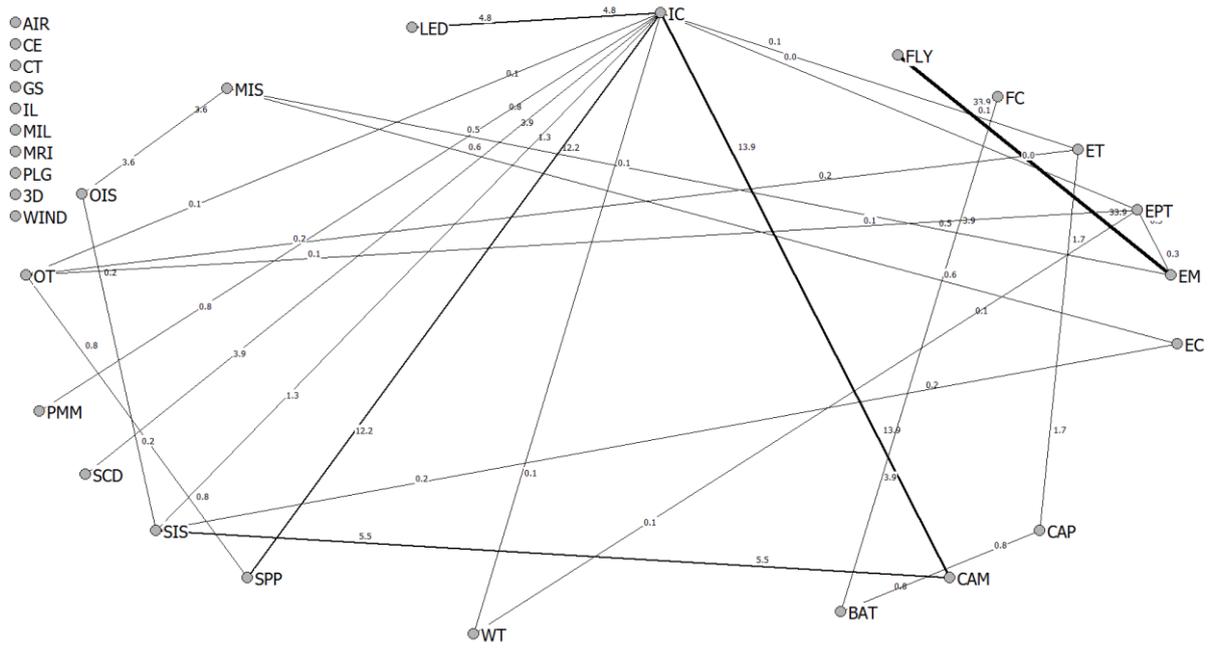

(a) Network at period 1

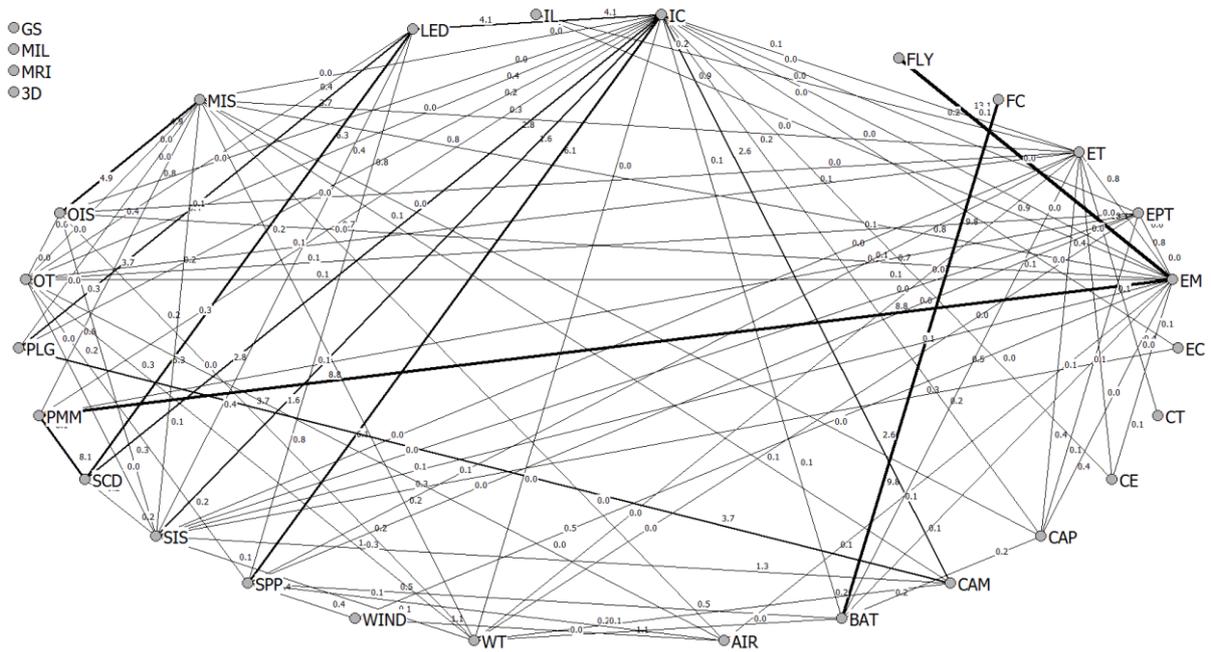

(b) Network at period 2

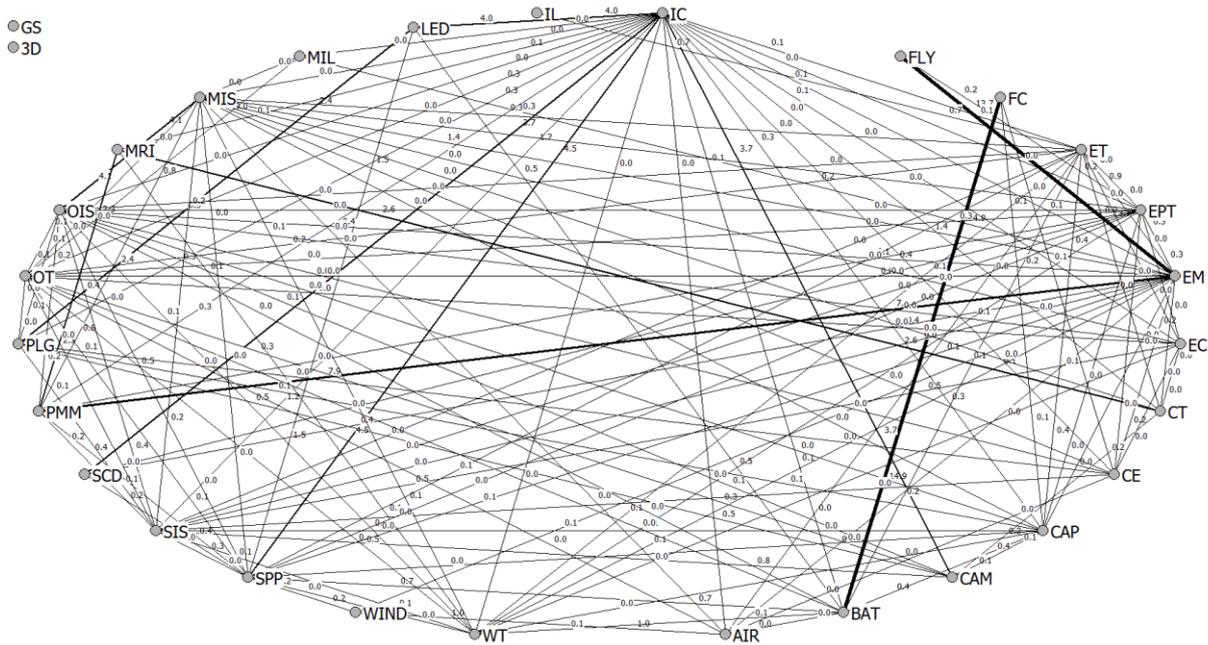

(c) Network at period 3

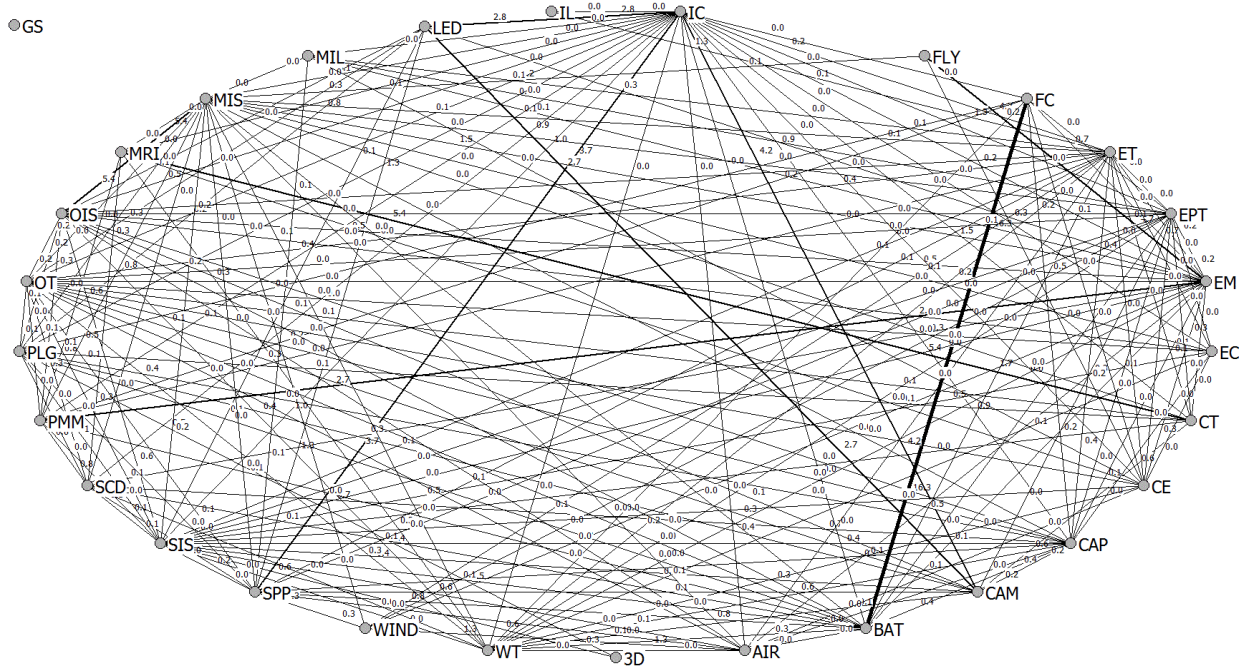

(d) Network at period 4

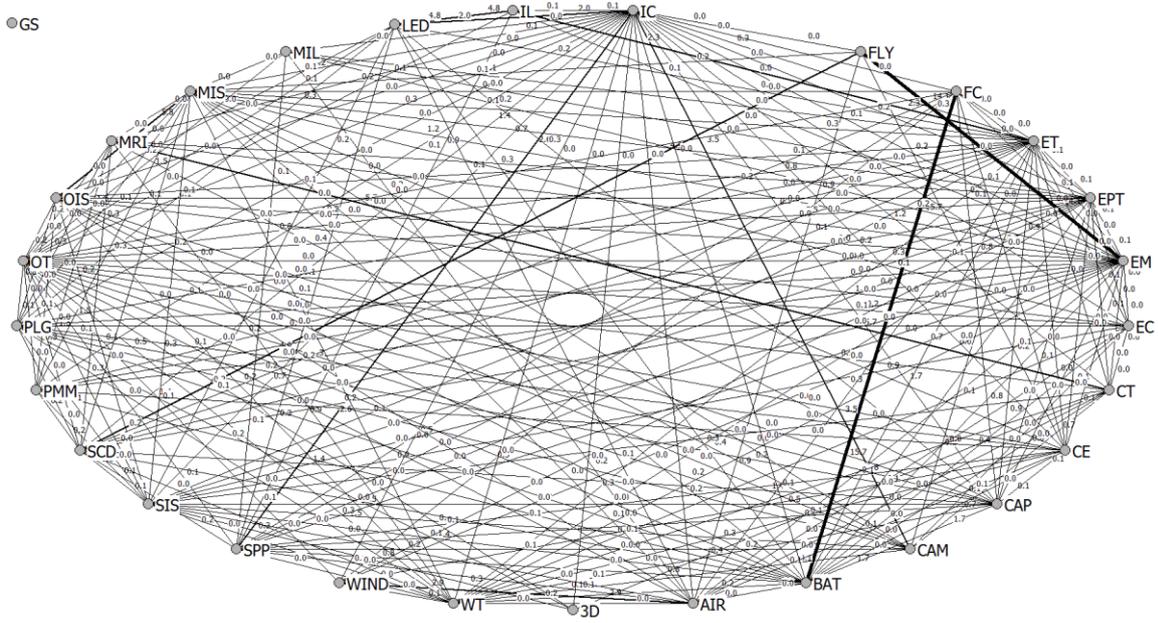

(e) Network at period 5

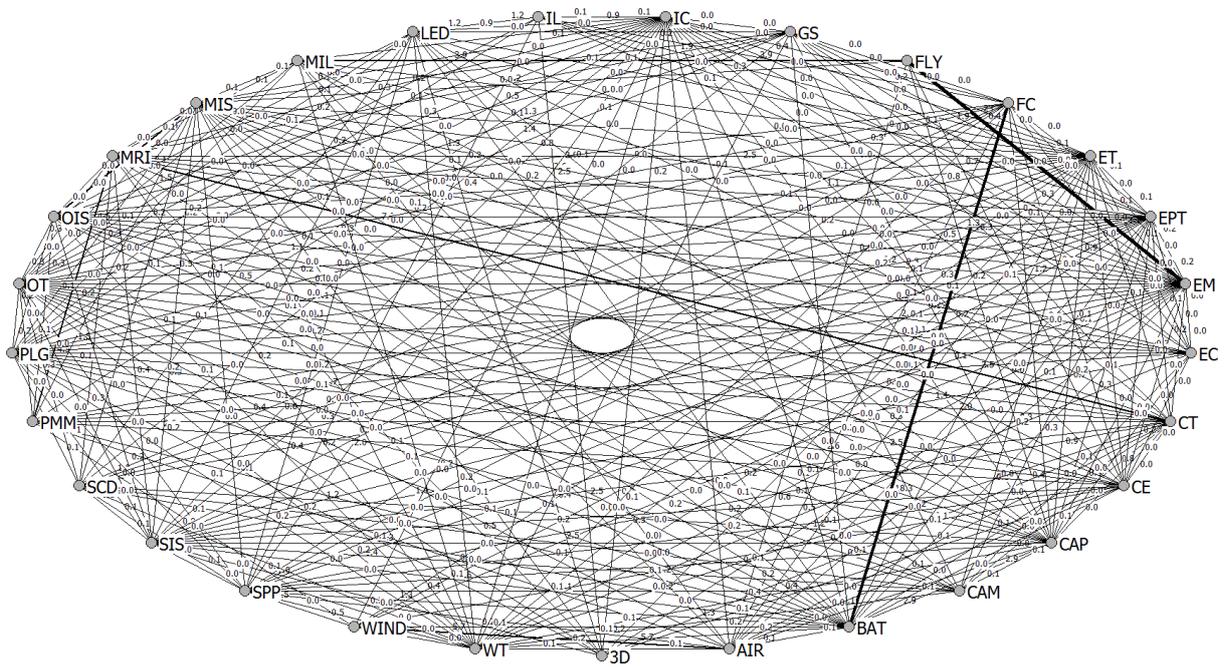

(f) Network at period 6

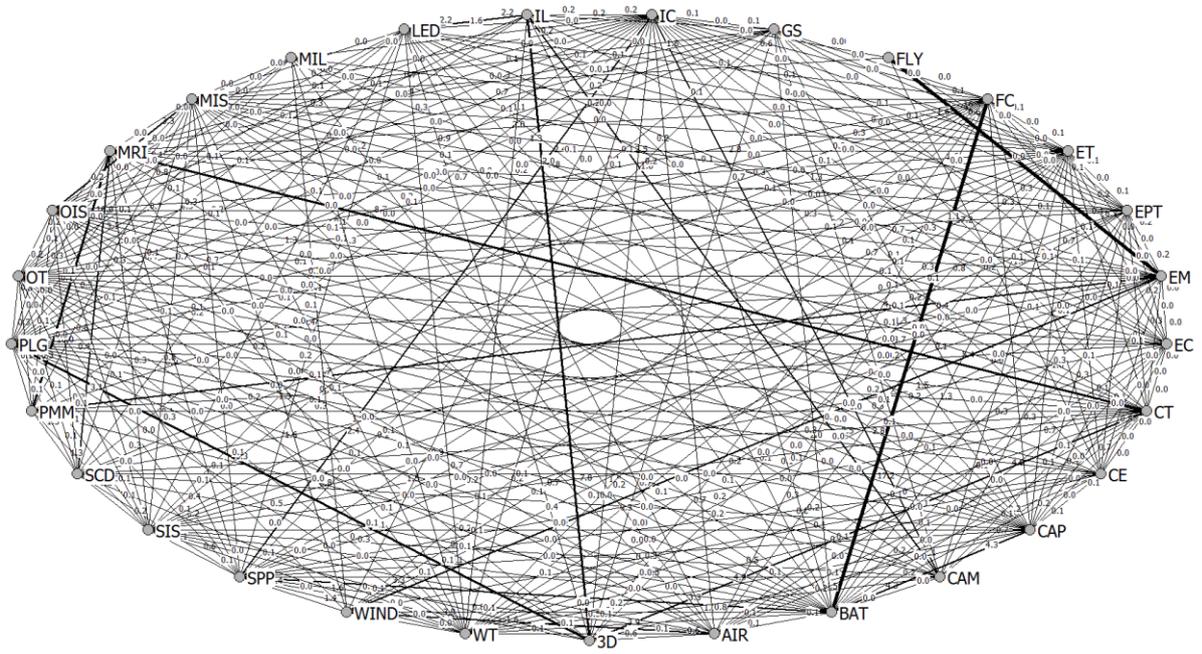

(g) Network at period 7

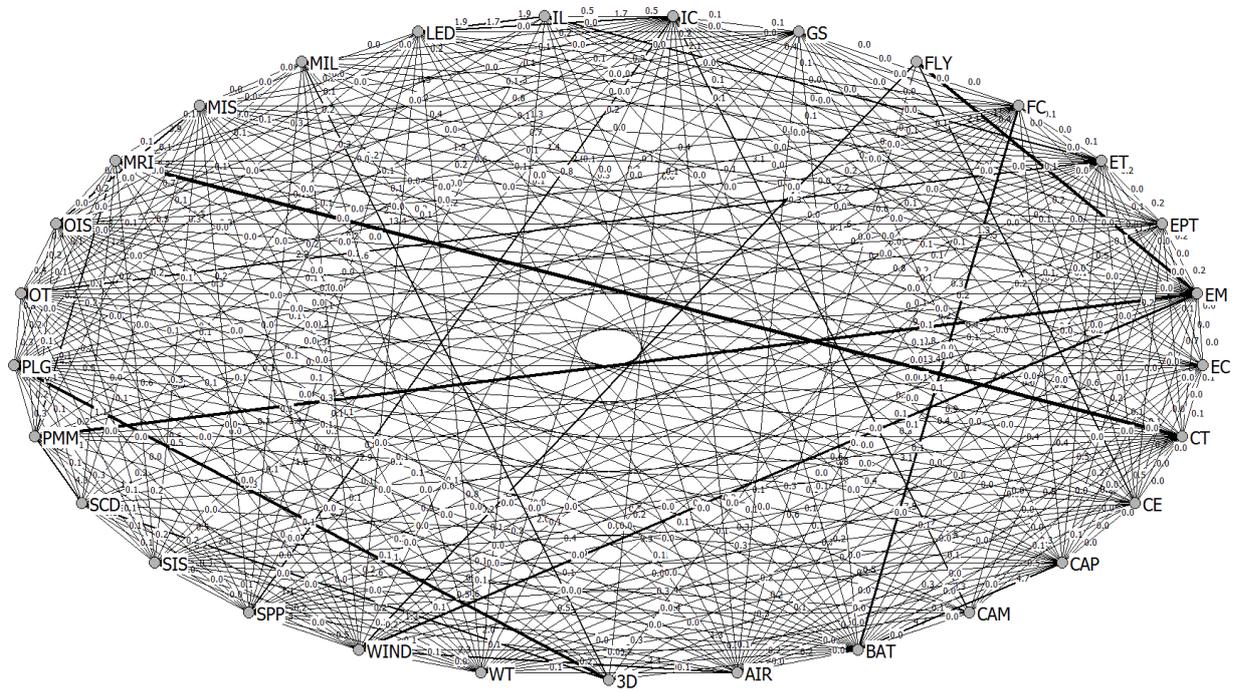

(a) Network at period 8

**Appendix 2. Cross-TD networks (strong links, testing networks)**

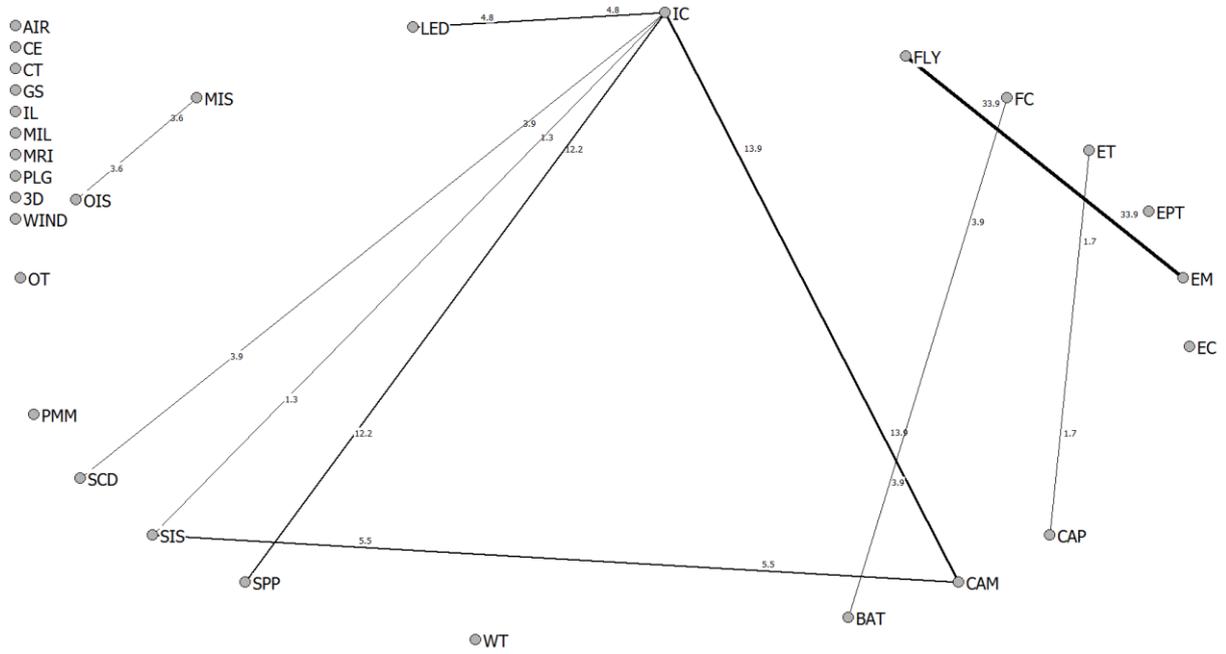

(a) Network at period 1 (cutoff =1)

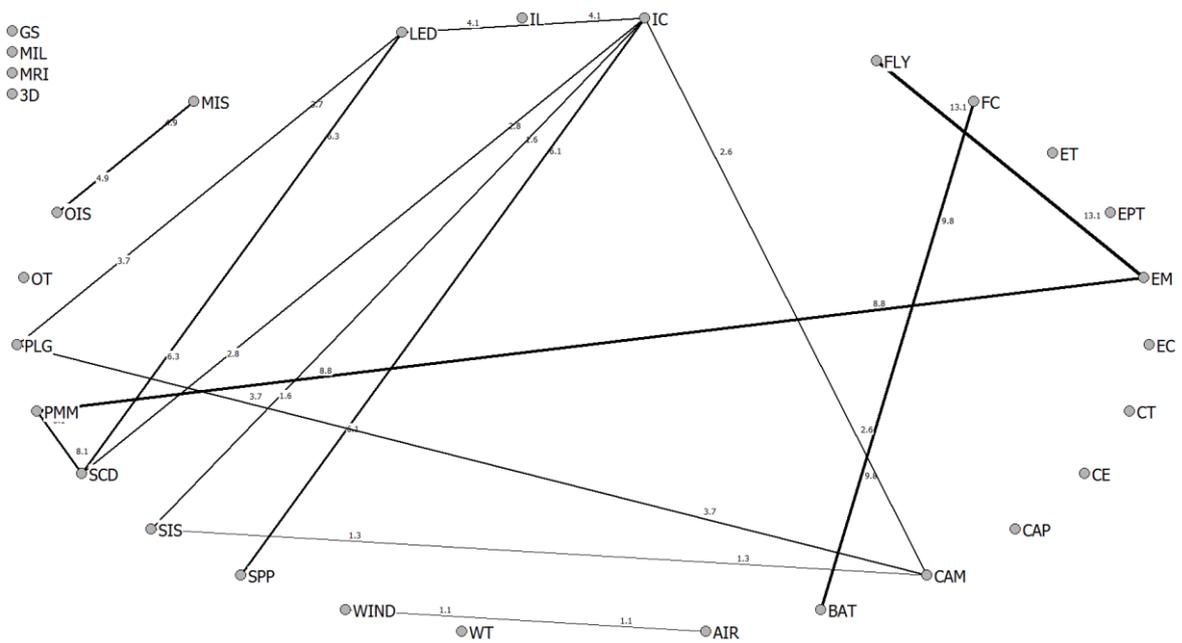

(b) Network at period 2 (cutoff =1)

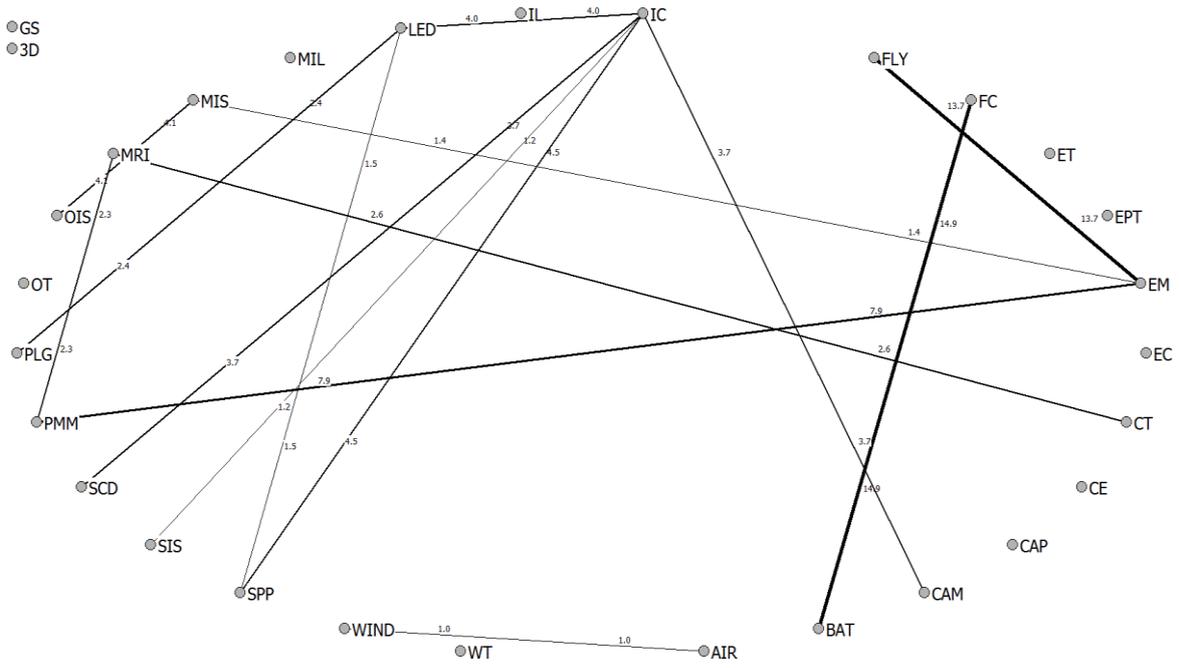

(c) Network at period 3 (cutoff =1)

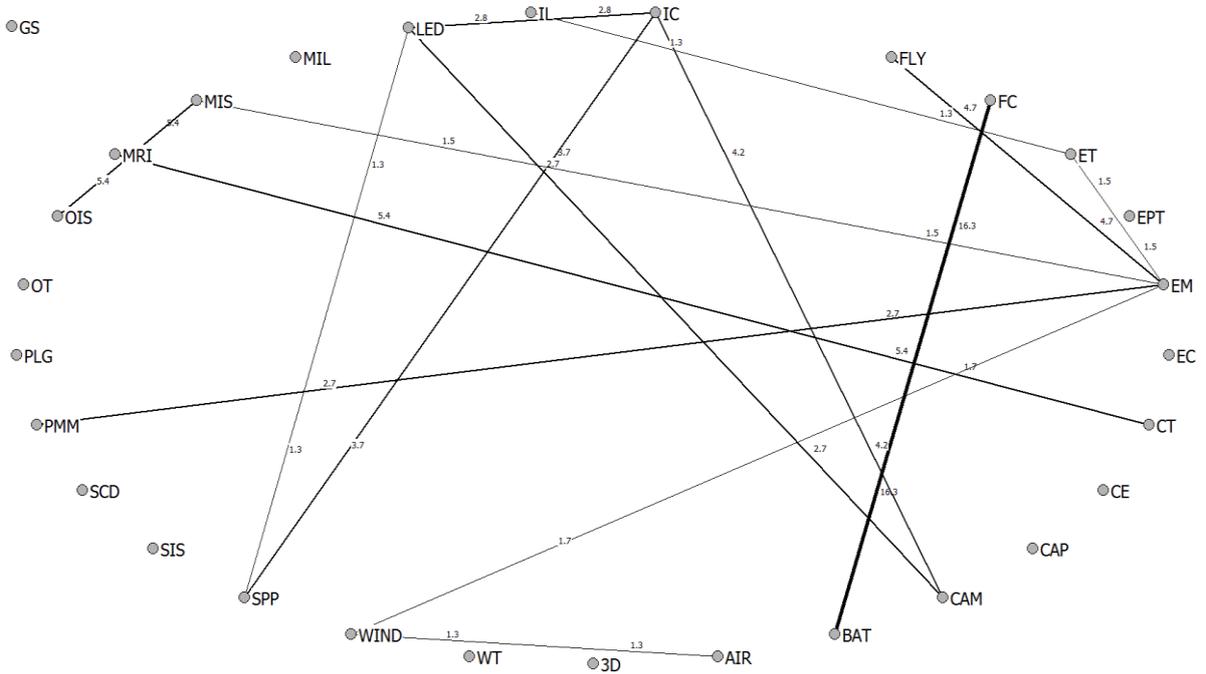

(d) Network at period 4 (cutoff =1)

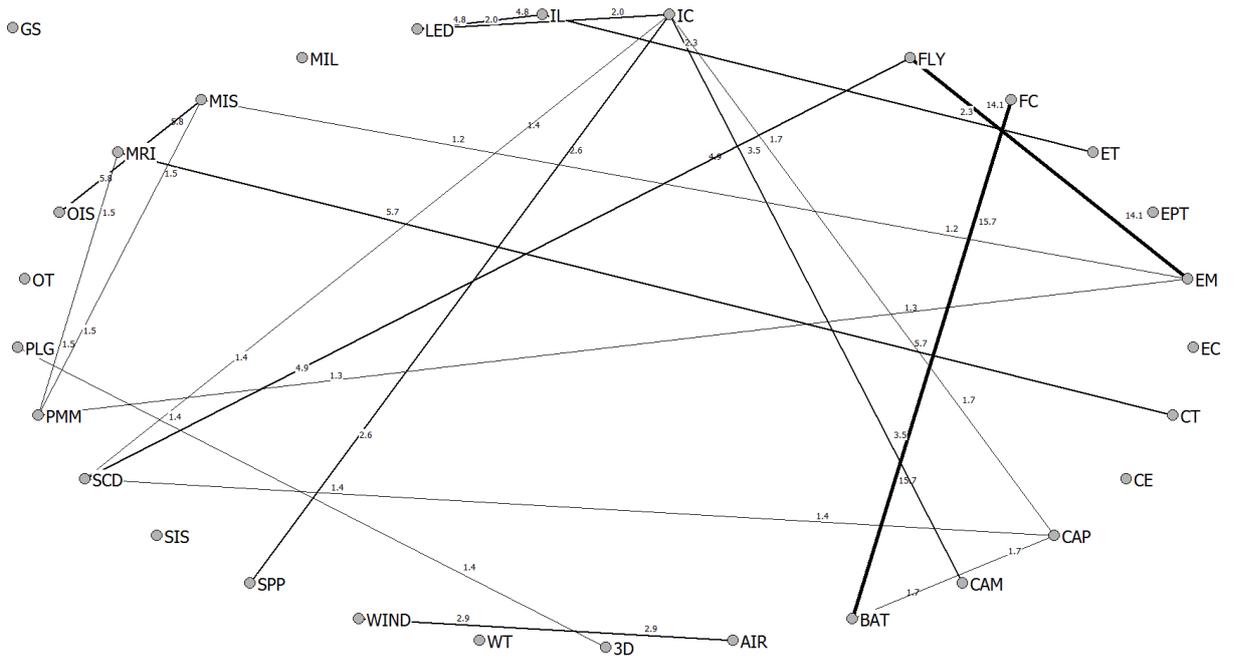

(e) Network at period 5 (cutoff =1)

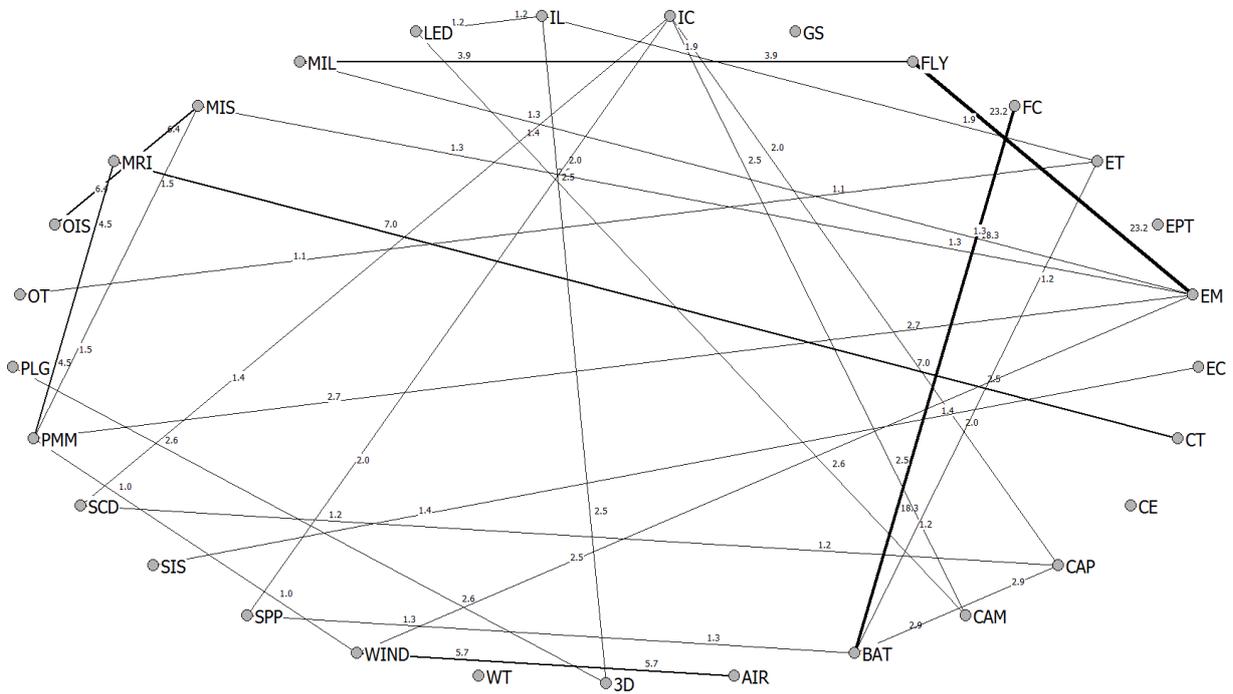

(f) Network at period 6 (cutoff =1)

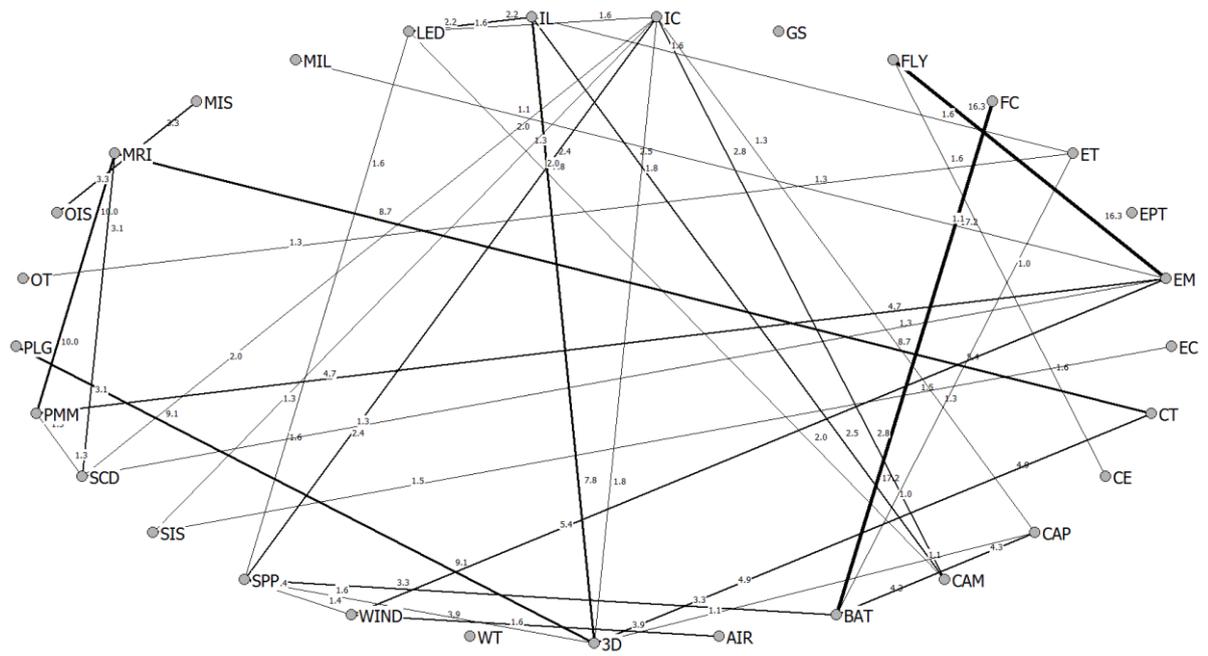

(g) Network at period 7 (cutoff =1)

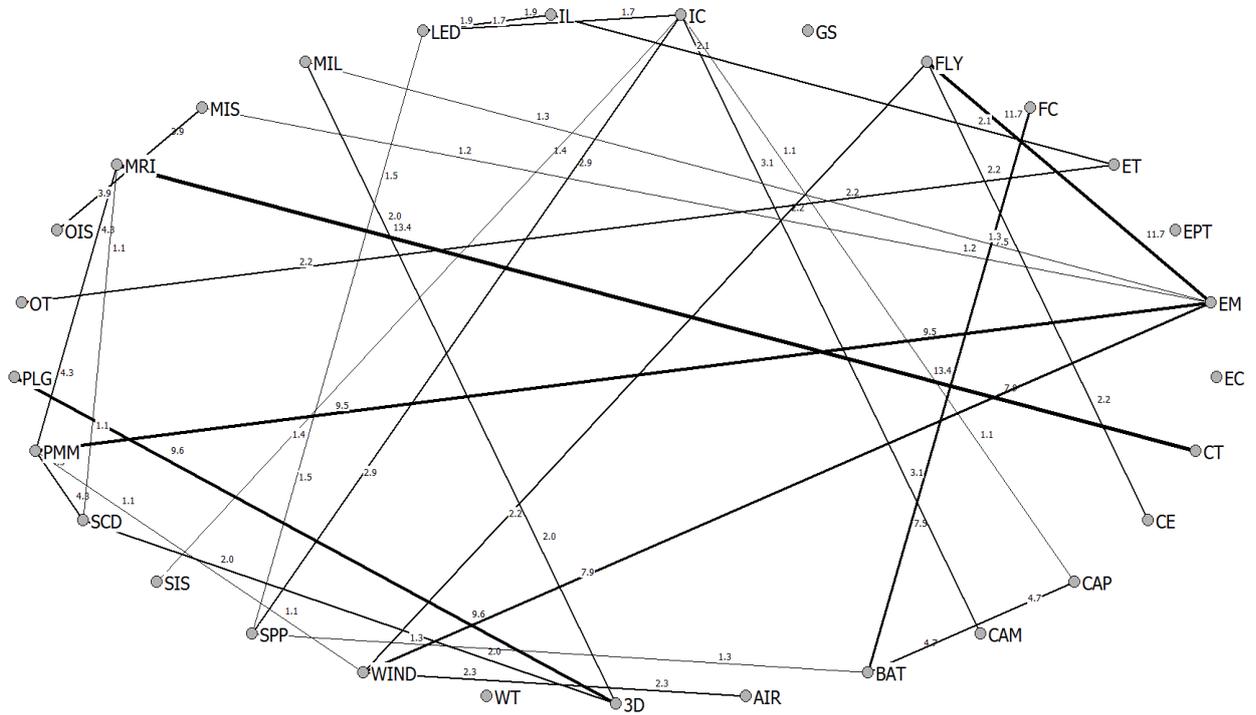

(h) Network at period 8 (cutoff =1)

**Appendix 3. Link prediction results: Number of predicted and actual links**

| | | T2 | T3 | T4 | T5 | T6 | T7 | T8 |
|---|---|---|---|---|---|---|---|---|
| Strong links | CommonNeighbors | 17 | 20 | 18 | 84 | 30 | 52 | 107 |
| | Jaccard | 23 | 53 | 54 | 103 | 91 | 151 | 142 |
| | Adamic-Adar | 19 | 24 | 27 | 99 | 65 | 104 | 136 |
| | ResourceAllocation | 35 | 75 | 124 | 155 | 193 | 253 | 251 |
| | PreferentialAttachment | 9 | 45 | 29 | 21 | 38 | 30 | 67 |
| | Katz ($\beta$=0.001) | 10 | 15 | 15 | 15 | 20 | 28 | 36 |
| | Katz ($\beta$=0.01) | 10 | 15 | 15 | 11 | 20 | 16 | 30 |
| | Katz ($\beta$=0.1) | 3 | 14 | 6 | 1 | 6 | 8 | 24 |
| | Katz ($\beta$=0.5) | 3 | 10 | 4 | 1 | 5 | 7 | 16 |
| | RootedPageRank ($\alpha$=0.01) | 23 | 39 | 38 | 55 | 55 | 79 | 78 |
| | RootedPageRank ($\alpha$=0.1) | 27 | 41 | 40 | 57 | 87 | 87 | 81 |
| | RootedPageRank ($\alpha$=0.5) | 55 | 56 | 63 | 77 | 87 | 146 | 137 |
| | RootedPageRank ($\alpha$=0.9) | 83 | 139 | 175 | 279 | 316 | 378 | 379 |
| | SimRank ($\gamma$=0.01) | 35 | 59 | 111 | 146 | 153 | 226 | 228 |
| | SimRank ($\gamma$=0.1) | 37 | 63 | 116 | 148 | 167 | 234 | 239 |
| | SimRank ($\gamma$=0.5) | 60 | 94 | 147 | 185 | 230 | 272 | 279 |
| | SimRank ($\gamma$=0.9) | 77 | 109 | 127 | 215 | 275 | 280 | 295 |
| | **Actual** | **15** | **15** | **15** | **20** | **28** | **36** | **29** |
| Weak links | CommonNeighbors | 53 | 231 | 304 | 287 | 348 | 354 | 299 |
| | Jaccard | 47 | 198 | 268 | 268 | 287 | 255 | 264 |
| | Adamic-Adar | 40 | 171 | 255 | 251 | 307 | 301 | 267 |
| | ResourceAllocation | 35 | 176 | 198 | 216 | 185 | 153 | 155 |
| | PreferentialAttachment | 141 | 181 | 252 | 283 | 290 | 348 | 330 |
| | Katz ($\beta$=0.001) | 9 | 33 | 29 | 29 | 87 | 95 | 231 |
| | Katz ($\beta$=0.01) | 9 | 33 | 29 | 33 | 87 | 107 | 237 |
| | Katz ($\beta$=0.1) | 16 | 34 | 38 | 43 | 101 | 115 | 243 |
| | Katz ($\beta$=0.5) | 16 | 38 | 40 | 43 | 102 | 116 | 251 |
| | RootedPageRank ($\alpha$=0.01) | 148 | 261 | 313 | 323 | 323 | 327 | 328 |
| | RootedPageRank ($\alpha$=0.1) | 144 | 259 | 311 | 321 | 291 | 319 | 325 |
| | RootedPageRank ($\alpha$=0.5) | 116 | 244 | 288 | 301 | 291 | 260 | 269 |
| | RootedPageRank ($\alpha$=0.9) | 88 | 161 | 176 | 99 | 62 | 28 | 27 |
| | SimRank ($\gamma$=0.01) | 136 | 241 | 240 | 232 | 225 | 180 | 178 |
| | SimRank ($\gamma$=0.1) | 134 | 237 | 235 | 230 | 211 | 172 | 167 |
| | SimRank ($\gamma$=0.5) | 111 | 206 | 204 | 193 | 148 | 134 | 127 |
| | SimRank ($\gamma$=0.9) | 94 | 191 | 224 | 163 | 103 | 126 | 111 |
| | **Actual** | **69** | **118** | **166** | **196** | **245** | **255** | **283** |

**Appendix 4. Link prediction results: Performance evaluation (Bolded entries are the best performances)**

| Accuracy | T2 | T3 | T4 | T5 | T6 | T7 | T8 | Ave. |
|---|---|---|---|---|---|---|---|---|
| CommonNeighbors | 0.941 | 0.934 | 0.938 | 0.793 | 0.877 | 0.823 | 0.744 | 0.864 |
| Jaccard | 0.926 | 0.852 | 0.850 | 0.741 | 0.751 | 0.638 | 0.648 | 0.772 |
| Adamic-Adar | 0.936 | 0.924 | 0.921 | 0.761 | 0.810 | 0.749 | 0.687 | 0.827 |
| ResourceAllocation | 0.902 | 0.823 | 0.717 | 0.633 | 0.559 | 0.451 | 0.434 | 0.645 |
| PreferentialAttachment | 0.956 | 0.882 | 0.911 | 0.904 | 0.852 | 0.857 | 0.813 | 0.882 |
| Katz($\beta$=0.001) | ***0.983*** | ***0.980*** | ***0.980*** | ***0.968*** | ***0.970*** | ***0.956*** | ***0.958*** | ***0.971*** |
| Katz($\beta$=0.01) | ***0.983*** | 0.975 | ***0.980*** | ***0.968*** | ***0.970*** | 0.941 | 0.953 | 0.967 |
| Katz($\beta$=0.1) | 0.970 | 0.958 | 0.973 | 0.953 | 0.936 | 0.916 | 0.929 | 0.948 |
| Katz($\beta$=0.5) | 0.970 | 0.963 | 0.973 | 0.953 | 0.938 | 0.919 | 0.938 | 0.951 |
| RootedPageRank($\alpha$=0.01) | 0.951 | 0.931 | 0.934 | 0.904 | 0.914 | 0.865 | 0.860 | 0.908 |
| RootedPageRank($\alpha$=0.1) | 0.941 | 0.926 | 0.929 | 0.899 | 0.840 | 0.845 | 0.852 | 0.890 |
| RootedPageRank($\alpha$=0.5) | 0.882 | 0.889 | 0.882 | 0.850 | 0.840 | 0.709 | 0.719 | 0.824 |
| RootedPageRank($\alpha$=0.9) | 0.813 | 0.685 | 0.606 | 0.362 | 0.291 | 0.158 | 0.138 | 0.436 |
| SimRank($\gamma$=0.01) | 0.902 | 0.857 | 0.749 | 0.650 | 0.653 | 0.507 | 0.490 | 0.687 |
| SimRank($\gamma$=0.1) | 0.897 | 0.847 | 0.737 | 0.645 | 0.618 | 0.488 | 0.463 | 0.671 |
| SimRank($\gamma$=0.5) | 0.850 | 0.786 | 0.660 | 0.559 | 0.473 | 0.404 | 0.370 | 0.586 |
| SimRank($\gamma$=0.9) | 0.803 | 0.749 | 0.709 | 0.500 | 0.377 | 0.389 | 0.335 | 0.552 |
| Precision | T2 | T3 | T4 | T5 | T6 | T7 | T8 | Ave. |
| CommonNeighbors | 0.235 | 0.200 | 0.222 | 0.119 | 0.133 | 0.154 | 0.150 | 0.173 |
| Jaccard | 0.174 | 0.076 | 0.074 | 0.087 | 0.099 | 0.133 | 0.099 | 0.106 |
| Adamic-Adar | 0.211 | 0.167 | 0.185 | 0.111 | 0.123 | 0.183 | 0.140 | 0.160 |
| ResourceAllocation | 0.143 | 0.120 | 0.097 | 0.084 | 0.109 | 0.130 | 0.100 | 0.112 |
| PreferentialAttachment | 0.333 | 0.133 | 0.138 | 0.048 | 0.079 | 0.133 | 0.149 | 0.145 |
| Katz($\beta$=0.001) | 0.900 | ***0.733*** | 0.733 | 0.733 | ***0.900*** | 0.821 | ***0.667*** | 0.784 |
| Katz($\beta$=0.01) | 0.900 | 0.667 | 0.733 | 0.818 | ***0.900*** | 0.875 | ***0.667*** | ***0.794*** |
| Katz($\beta$=0.1) | ***1.000*** | 0.429 | 0.833 | ***1.000*** | 0.667 | 0.625 | 0.500 | 0.722 |
| Katz($\beta$=0.5) | ***1.000*** | 0.500 | ***1.000*** | ***1.000*** | 0.800 | 0.714 | 0.625 | 0.806 |
| RootedPageRank($\alpha$=0.01) | 0.391 | 0.333 | 0.342 | 0.327 | 0.436 | 0.380 | 0.321 | 0.362 |
| RootedPageRank($\alpha$=0.1) | 0.333 | 0.317 | 0.325 | 0.316 | 0.287 | 0.345 | 0.309 | 0.319 |
| RootedPageRank($\alpha$=0.5) | 0.200 | 0.232 | 0.238 | 0.234 | 0.287 | 0.219 | 0.190 | 0.229 |
| RootedPageRank($\alpha$=0.9) | 0.133 | 0.094 | 0.086 | 0.072 | 0.089 | 0.095 | 0.077 | 0.092 |
| SimRank($\gamma$=0.01) | 0.143 | 0.136 | 0.108 | 0.082 | 0.131 | 0.137 | 0.110 | 0.121 |
| SimRank($\gamma$=0.1) | 0.135 | 0.127 | 0.103 | 0.081 | 0.120 | 0.133 | 0.105 | 0.115 |
| SimRank($\gamma$=0.5) | 0.117 | 0.117 | 0.082 | 0.070 | 0.096 | 0.121 | 0.093 | 0.099 |
| SimRank($\gamma$=0.9) | 0.078 | 0.101 | 0.095 | 0.074 | 0.091 | 0.121 | 0.092 | 0.093 |
| Recall | T2 | T3 | T4 | T5 | T6 | T7 | T8 | Ave. |
| CommonNeighbors | 0.267 | 0.267 | 0.267 | 0.500 | 0.143 | 0.222 | 0.552 | 0.317 |
| Jaccard | 0.267 | 0.267 | 0.267 | 0.450 | 0.321 | 0.556 | 0.483 | 0.373 |
| Adamic-Adar | 0.267 | 0.267 | 0.333 | 0.550 | 0.286 | 0.528 | 0.655 | 0.412 |
| ResourceAllocation | 0.333 | 0.600 | 0.800 | 0.650 | 0.750 | 0.917 | 0.862 | 0.702 |
| PreferentialAttachment | 0.200 | 0.400 | 0.267 | 0.050 | 0.107 | 0.111 | 0.345 | 0.211 |
| Katz($\beta$=0.001) | 0.600 | 0.733 | 0.733 | 0.550 | 0.643 | 0.639 | 0.828 | 0.675 |
| Katz(*β*=0.01) | 0.600 | 0.667 | 0.733 | 0.450 | 0.643 | 0.389 | 0.690 | 0.596 |
| Katz(*β* =0.1) | 0.200 | 0.400 | 0.333 | 0.050 | 0.143 | 0.139 | 0.414 | 0.240 |
| Katz(*β* =0.5) | 0.200 | 0.333 | 0.267 | 0.050 | 0.143 | 0.139 | 0.345 | 0.211 |
| RootedPageRank(*α*=0.01) | 0.600 | ***0.867*** | 0.867 | 0.900 | 0.857 | 0.833 | 0.862 | 0.827 |
| RootedPageRank(*α*=0.1) | 0.600 | ***0.867*** | 0.867 | 0.900 | 0.893 | 0.833 | 0.862 | 0.832 |
| RootedPageRank(*α*=0.5) | ***0.733*** | ***0.867*** | ***1.000*** | 0.900 | 0.893 | 0.889 | 0.897 | 0.883 |
| RootedPageRank(*α*=0.9) | ***0.733*** | ***0.867*** | ***1.000*** | ***1.000*** | ***1.000*** | ***1.000*** | ***1.000*** | ***0.943*** |
| SimRank(*γ*=0.01) | 0.333 | 0.533 | 0.800 | 0.600 | 0.714 | 0.861 | 0.862 | 0.672 |
| SimRank(*γ*=0.1) | 0.333 | 0.533 | 0.800 | 0.600 | 0.714 | 0.861 | 0.862 | 0.672 |
| SimRank(*γ*=0.5) | 0.467 | 0.733 | 0.800 | 0.650 | 0.786 | 0.917 | 0.897 | 0.750 |
| SimRank(*γ*=0.9) | 0.400 | 0.733 | 0.800 | 0.800 | 0.893 | 0.944 | 0.931 | 0.786 |
| F-score | T2 | T3 | T4 | T5 | T6 | T7 | T8 | Ave. |
| CommonNeighbors | 0.250 | 0.229 | 0.242 | 0.192 | 0.138 | 0.182 | 0.235 | 0.210 |
| Jaccard | 0.211 | 0.118 | 0.116 | 0.146 | 0.151 | 0.214 | 0.164 | 0.160 |
| Adamic-Adar | 0.235 | 0.205 | 0.238 | 0.185 | 0.172 | 0.271 | 0.230 | 0.220 |
| ResourceAllocation | 0.200 | 0.200 | 0.173 | 0.149 | 0.190 | 0.228 | 0.179 | 0.188 |
| PreferentialAttachment | 0.250 | 0.200 | 0.182 | 0.049 | 0.091 | 0.121 | 0.208 | 0.157 |
| Katz(*β*=0.001) | ***0.720*** | ***0.733*** | ***0.733*** | ***0.629*** | ***0.750*** | ***0.719*** | ***0.739*** | ***0.718*** |
| Katz(*β*=0.01) | ***0.720*** | 0.667 | ***0.733*** | 0.581 | ***0.750*** | 0.539 | 0.678 | 0.667 |
| Katz(*β* =0.1) | 0.333 | 0.414 | 0.476 | 0.095 | 0.235 | 0.227 | 0.453 | 0.319 |
| Katz(*β* =0.5) | 0.333 | 0.400 | 0.421 | 0.095 | 0.242 | 0.233 | 0.444 | 0.310 |
| RootedPageRank(*α*=0.01) | 0.474 | 0.482 | 0.491 | 0.480 | 0.578 | 0.522 | 0.467 | 0.499 |
| RootedPageRank(*α*=0.1) | 0.429 | 0.464 | 0.473 | 0.468 | 0.435 | 0.488 | 0.455 | 0.459 |
| RootedPageRank(*α*=0.5) | 0.314 | 0.366 | 0.385 | 0.371 | 0.435 | 0.352 | 0.313 | 0.362 |
| RootedPageRank(*α*=0.9) | 0.225 | 0.169 | 0.158 | 0.134 | 0.163 | 0.174 | 0.142 | 0.166 |
| SimRank(*γ*=0.01) | 0.200 | 0.216 | 0.191 | 0.145 | 0.221 | 0.237 | 0.195 | 0.201 |
| SimRank(*γ*=0.1) | 0.192 | 0.205 | 0.183 | 0.143 | 0.205 | 0.230 | 0.187 | 0.192 |
| SimRank(*γ*=0.5) | 0.187 | 0.202 | 0.148 | 0.127 | 0.171 | 0.214 | 0.169 | 0.174 |
| SimRank(*γ*=0.9) | 0.130 | 0.177 | 0.169 | 0.136 | 0.165 | 0.215 | 0.167 | 0.166 |